\documentclass[12pt]{article}
\pdfoutput=1
\usepackage{graphicx}
\usepackage{dsfont}
\usepackage{amssymb}
\usepackage{amsmath}
\usepackage{mathrsfs}

\begin{document}

\title{Fiber-optical analogue of the event horizon: Appendices}
\author{Thomas G.\ Philbin$^{1,2}$,
Chris Kuklewicz$^{1}$,
Scott Robertson$^{1}$,
Stephen Hill$^{1}$,\\
Friedrich K\"onig$^{1}$,
and
Ulf Leonhardt$^{1}$\\
$^1$School of Physics and Astronomy,\\ University of St Andrews,
North Haugh, St Andrews, Fife, KY16 9SS, UK\\
$^2$Max Planck Research Group of Optics, Information and Photonics,\\
G\"unther-Scharowsky-Str.\ 1, Bau 24, D-91058 Erlangen, Germany
}
\date{\today}
\maketitle
\begin{abstract}
We explain the theory behind our
fiber-optical analogue of the event horizon
and present the experiment in detail.
\end{abstract}

\maketitle

\newpage

\begin{appendix}

\renewcommand{\theequation}{A\arabic{equation}}
\setcounter{equation}{0}
\renewcommand{\thefigure}{A\arabic{figure}}
\setcounter{figure}{0}

\section{Theory}

In this appendix we describe the theory behind our fiber-optical
analogue of the event horizon \cite{Philbin}. 
After a brief summary of Nonlinear
Fiber Optics we show how an optical pulse establishes a moving
medium and how this medium corresponds to a space-time geometry.
This point of view is relatively unusual in the fiber optics
community, but, as we demonstrate, it is completely consistent
with the established knowledge of this field \cite{Agrawal}. We
then take advantage of our approach in quantizing the
electromagnetic field of a probe in the presence of a pulse. We
describe the classical and quantum physics of fiber-optical
horizons, explaining the classical frequency shifting and the
quantum Hawking effect with as few assumptions on the physics and
the prior knowledge of the reader as possible without going into
excessive detail.

\subsection{Kerr nonlinearity}

Optical fibers
\cite{Agrawal,Russells} are metamaterials ---
materials with optical properties that are dominated by their structure ---
for the following reason:
an optical fiber confines light to narrow transversal regions,
usually at the core of the fiber.
The electromagnetic field in the transversal plane
depends on the frequency and polarization of the light
in relation to the transversal structure of the fiber.
In this way \cite{Agrawal}, the waveguide gives rise to an effective
dispersion \cite{BornWolf} and birefringence \cite{BornWolf}
that supersedes the natural optical dispersion of silica glass.
The confinement of the light in the fiber also enhances
the intensity over the fiber length
such that the natural optical nonlinearity
of glass becomes relevant, in particular the
Kerr and Raman nonlinearity caused, respectively,
by the electronic and molecular
optical response of glass \cite{Agrawal}.
For making artificial event horizons,
one can exploit the custom-designed dispersion
and birefringence and the enhanced nonlinearity
of optical fibers,
in particular of microstructured fibers
(also called photonic-crystal fibers) \cite{Russells}.

Let us first describe the optical properties of glass.
In the laboratory frame,
$\mathbf{r}=(x,y,z)$ describes the position in
Cartesian coordinates and $t$ the time.
We use the operator $\nabla$ to denote spatial derivatives and
$\partial_t$ for time derivatives.
We characterize the intrinsic linear optical response of the glass
fiber by the susceptibility profile
$\chi_{_g}(t,\mathbf{r})$
and denote the nonlinear polarization vector of the medium
by $\mathbf{P}$.
The vector $\mathbf{E}$ of the electric field strength
obeys the wave equation \cite{LL8} in SI units,
%%%%%%%%%%%%%%%%%%
\begin{equation}
\nabla\times(\nabla\times\mathbf{E}) +
\frac{\partial_t^2}{c^2}\left(
\mathbf{E} +
\int_{-\infty}^t \chi_{_g}(t-t')\,
\mathbf{E}(t')\,dt'
+\frac{\mathbf{P}}{\varepsilon_0}
\right)
=0
\end{equation}
%%%%%%%%%%%%%%%%%%
with $\varepsilon_0$
being the electric permeability of the vacuum.
We assume a lossless medium where $\chi_g$ is real.
In isotropic materials such as glass,
no second-order nonlinearity exists on symmetry grounds
\cite{NLO}.
The lowest-order nonlinearity is proportional
to the response from the
products of three electric-field components.
Such an effect,
called Kerr nonlinearity \cite{Agrawal}
is generated by the electronic excitations of glass
on a time scale comparable to the atomic size divided
by the speed of light.
Assuming instantaneous response, the
nonlinear polarization is proportional
to a cubic form of the field strength.
The effect of optical nonlinearities
strongly depends on the frequency matching
of the field components involved \cite{NLO}.
Consider instead of the full electric field $\mathbf{E}$
either the positive or negative frequency component
oscillating in the optical spectral range.
To keep the notation simple, we describe these components
by the symbol $\mathbf{E}$ as well.
The nonlinear polarization of the medium is only effective
if $\mathbf{P}$ oscillates in a similar spectral range
as one of the frequency components of $\mathbf{E}$.
Consequently,
$\mathbf{P}$ combines  two electric-field components
$E_l$ and one complex conjugate $E_m^*$.
In isotropic materials such as glass, $\mathbf{P}$
is further restricted:
isotropy implies for the $\mathbf{P}$ components
\cite{NLO}
%%%%%%%%%%%%%%%%%%
\begin{equation}
\label{eq:kerrpol}
P_m = \frac{2\varepsilon_0}{3}
\sum_l \bigg(
\kappa_1 E_m E_l E_l^* +
\kappa_2 E_l E_m E_l^* +
\kappa_3 E_l E_l E_m^*
\bigg)
\end{equation}
%%%%%%%%%%%%%%%%%%
where the $\kappa$ denote the material constants
of the Kerr nonlinearity.
They correspond to the $\chi^{(3)}$
coefficients in nonlinear optics \cite{NLO}.
In silica,
the $\kappa$ constants are identical,
 to a very good approximation,
%%%%%%%%%%%%%%%%%%
\begin{equation}
\label{eq:kappa}
\kappa_1 =
\kappa_2 =
\kappa_3 =
\kappa \,.
\end{equation}
%%%%%%%%%%%%%%%%%%
In fibers \cite{Agrawal},
the electric field has two transversal components
that we denote by $\pm$.
Equations (\ref{eq:kerrpol}) and (\ref{eq:kappa}) give
%%%%%%%%%%%%%%%%%%
\begin{equation}
\label{eq:pol}
P_\pm = 2\varepsilon_0\kappa
\bigg(
|E_\pm|^2E_\pm + \frac{2}{3}|E_\mp|^2E_\pm +
\frac{1}{3}E_\mp^2 E_\pm^*
\bigg) .
\end{equation}
%%%%%%%%%%%%%%%%%%
The first term describes the Self Phase Modulation,
the second term the Cross Phase Modulation
and the third usually is called Four Wave Mixing
\cite{Agrawal}.
The Self Phase Modulation does not act across polarizations,
whereas the other two terms correspond
to polarization interactions.
For equally polarized fields that carry two distinct
frequency bands,
the Kerr effect acts across these frequencies,
because, writing
%%%%%%%%%%%%%%%%%%
\begin{equation}
E_\pm = E_a e^{-i\omega_at} + E_b e^{-i\omega_bt}
\end{equation}
%%%%%%%%%%%%%%%%%%
we obtain
%%%%%%%%%%%%%%%%%%
\begin{equation}
|E_\pm|^2 E_\pm = (|E_a|^2+2|E_b|^2)E_a e^{-i\omega_at}
+  (|E_b|^2+2|E_a|^2)E_b e^{-i\omega_bt} \,.
\end{equation}
%%%%%%%%%%%%%%%%%%
The contribution proportional to $|E_b|^2E_a$ or $|E_a|^2E_b$
is also called Cross Phase Modulation \cite{Agrawal}.
We see that the coupling strength of the
Cross Phase Modulation between polarizations
is $2/3$ of the Self Phase Modulation
and the Cross Phase Modulation within one polarization
is twice as strong as the Self Phase Modulation \cite{Agrawal}.

Another cubic nonlinear effect
caused by molecular excitations
with longer response time,
called Stimulated Raman Scattering \cite{Agrawal},
contributes
to the material polarization
for pulses below $100\mathrm{fs}$ duration.
Stimulated Raman Scattering
is important for forming the shape of
ultrashort pulses \cite{Agrawal},
but, it does not act across polarizations
nor significantly different frequency bands,
the cases we are interested in;
hence we ignore the Raman effect here.

\subsection{Waveguides}

Consider the effect of the waveguide on the light
confined in the $x$ and $y$ direction and propagating
along the fiber in the $z$ direction.
We assume that the fiber is homogeneous in $z$ and
infinitely long,
with the
Fourier-transformed susceptibility
%%%%%%%%%%%%%%%%%%
\begin{equation}
\widetilde{\chi}_{_g} =
\widetilde{\chi}_{_g}(\omega,x,y)
\,.
\end{equation}
%%%%%%%%%%%%%%%%%%
We represent
the Fourier-transformed field strengths as
%%%%%%%%%%%%%%%%%%
\begin{equation}
\widetilde{\mathbf{E}}(\omega,\mathbf{r}) =
\widetilde{E}_\pm(\omega,z)\,
\mathbf{U}_\pm(\omega,x,y)
\end{equation}
%%%%%%%%%%%%%%%%%%
and require that the fiber modes
$\mathbf{U}_\pm$ are eigenfunctions
of the transversal part of
the wave equation for monochromatic light
with eigenvalues $\beta^2_\pm(\omega)$,
%%%%%%%%%%%%%%%%%%
\begin{equation}
\label{eq:upm}
\left(
-\nabla\times(\nabla\times\mathbf{U}_\pm)
+ (1+\widetilde{\chi}_{_g}) \frac{\omega^2}{c^2}\mathbf{U}_\pm
\right)
=\beta_\pm^2(\omega)\mathbf{U}_\pm
\,\,.
\end{equation}
%%%%%%%%%%%%%%%%%%
For single-mode fibers \cite{Agrawal},
only one eigenvalue $\beta_\pm^2(\omega)$ exists
for each optical polarization $\pm$ and frequency $\omega$
(within a limited frequency range).
We normalize the mode functions $\mathbf{U}_\pm$ such that
the integral of $|\mathbf{U}_\pm(\omega,x,y)|^2$ over
the $(x,y)$ plane is unity.
The optical nonlinearity acts predominantly
within a narrow spatial region near the fiber core at
$(x,y)=(0,0)$. Therefore,
we obtain, to a very good approximation,
%%%%%%%%%%%%%%%%%%
\begin{equation}
\label{eq:nlwave}
\big[
\partial_z^2 + \beta_\pm^2(i\partial_t)
\big] E_\pm
= \frac{\partial_t^2 P_\pm}{\varepsilon_0 c^2}
\end{equation}
%%%%%%%%%%%%%%%%%%
with the nonlinear polarization (\ref{eq:pol}).
Here we understand $\kappa$ as being averaged
over the transversal modes (usually causing
almost no difference between the two
optical polarizations $\pm$).

The eigenvalues $\beta^2_\pm(\omega)$
of the transversal modes set the effective
refractive indices $n_\pm$ of the
fiber for light pulses $E_\pm(t,z)$ defined by the relation
%%%%%%%%%%%%%%%%%%
\begin{equation}
\label{eq:beta}
\beta_\pm = \frac{n_\pm}{c}\,\omega \,.
\end{equation}
%%%%%%%%%%%%%%%%%%
As a well-known consequence of causality \cite{Toll}
the refractive index must be analytic
on the upper half plane of complex $\omega$.
In the absence of losses within the frequency range
we are considering,
the Fourier-transformed $\widetilde{\chi}_g(\omega)$
in the longitudinal mode equation (\ref{eq:upm})
is real on the real axis
and the longitudinal mode equation (\ref{eq:upm})
is Hermitian and positive.
Consequently, $n_\pm$ is real on the real axis.
Furthermore, since the linear susceptibility $\chi_g(t)$ is real,
$\widetilde{\chi}_g(\omega)$ must be an even function of
$\omega$, which implies that 
$n_\pm^2$ and $\beta^2$ are even functions
of the frequency $\omega$.

When the dielectric structure of the fiber
varies over the scale of an optical wavelength,
the polarization of light becomes an important issue
and the refractive indices differ for
$E_+(t,z)$ and $E_-(t,z)$,
causing birefingence \cite{BornWolf}.
As we have seen,
microstructured or photonic-crystal fibers \cite{Russells}
allow for some freedom in tailoring
the dispersion, nonlinearity, and birefringence
for specific applications.

\subsection{Effective moving medium}

In our case, an intense ultrashort optical pulse interacts with
a weak probe field. This probe may be caused by an incident
continuous wave of light or
by the vacuum fluctuations
of the electromagnetic field itself \cite{MilonniQV}.
The vacuum fluctuations are carried by modes that
behave as weak classical light fields as well.
The pulse is polarized along one of the eigen-polarizations
of the fiber; the probe field may be co- or cross polarized.
Due to Stimulated Raman Scattering \cite{Agrawal}
or the formation of optical shocks \cite{Agrawal}
the pulse will change its shape and velocity;
but we assume that the intensity profile $I(z,t)$ of the pulse
uniformly moves with constant velocity $u$
during the interaction with the probe.
Since the probe field is weak or in the vacuum state
we can safely neglect the backaction onto the pulse
and the nonlinear self interaction of the probe.
Since the intensity profile of the pulse is assumed to be fixed
and given, we focus attention on the probe field.
We describe the probe by the
corresponding component $A$ of the vector potential
that generates the electric field $E$ and
the magnetic field $B$, with
 %%%%%%%%%%%%%%%%%%
\begin{equation}
E=-\partial_t A \,,
\quad
B=\partial_z A \,.
\end{equation}
%%%%%%%%%%%%%%%%%%
The probe field obeys the wave equation
 %%%%%%%%%%%%%%%%%%
\begin{equation}
\label{eq:wave0}
\big(c^2\partial_z^2 + c^2\beta^2(i\partial_t)
-\partial_t\chi\partial_t \big) A = 0 \,,\quad
\chi \propto I(z,t)
\end{equation}
%%%%%%%%%%%%%%%%%%
where $\chi$ denotes the susceptibility due to the
Kerr effect of the pulse on the probe.
We take the $\beta$ of Eq.\ (\ref{eq:beta})
that corresponds to the probe polarization
and denote the effective refractive index by $n_0$.
Equation (\ref{eq:wave0}) shows that
the pulse indeed establishes
an effective moving medium
\cite{LeoReview}.
It is advantageous \cite{Agrawal} to use as coordinates
the retarded time $\tau$ and the propagation time $\zeta$
defined as
 %%%%%%%%%%%%%%%%%%
\begin{equation}
\label{eq:frame}
\tau=t- \frac{z}{u}\,,\quad \zeta=\frac{z}{u} \,,
\end{equation}
%%%%%%%%%%%%%%%%%%
because in this case the properties of the effective medium
depend only on $\tau$.
Here $\tau$ plays the role of space and $\zeta$ of time.
In this co-moving frame we replace
in the wave equation (\ref{eq:wave0}) the
$z$ and $t$ derivatives by
 %%%%%%%%%%%%%%%%%%
\begin{equation}
\label{eq:derivatives}
\partial_t=\partial_{\tau} \,,\quad
\partial_z =\frac{1}{u}(\partial_\zeta - \partial_\tau) \,,
\end{equation}
%%%%%%%%%%%%%%%%%%
and obtain
 %%%%%%%%%%%%%%%%%%
\begin{equation}
\label{eq:wave}
(\partial_\zeta-\partial_\tau)^2 A =
\partial_\tau \frac{u^2}{c^2}n^2\partial_\tau A \,,
\end{equation}
%%%%%%%%%%%%%%%%%%
where the total refractive index $n$ consists of the
effective linear index $n_0$ and the
contribution due the Kerr effect of the pulse,
%%%%%%%%%%%%%%%%%%
\begin{equation}
\label{eq:n}
n^2 = n_0^2 +\chi \,.
\end{equation}
%%%%%%%%%%%%%%%%%%
Since $\chi\ll n_0$ we approximate
%%%%%%%%%%%%%%%%%%
\begin{equation}
\label{eq:approxn}
n \approx n_0 +\delta n \,,\quad \delta n = \frac{\chi}{2n_0}  \,,
\end{equation}
%%%%%%%%%%%%%%%%%%
where we can ignore the frequency dependance of $n_0$ in $\chi/(2n_0)$,
which gives Eq.\ (1) of our paper \cite{Philbin}.
Note that Eq.\ (\ref{eq:frame}) does not describe a Lorentz
transformation, but the $\tau$ and $\zeta$ are of course
perfectly valid coordinates; they simply do not
belong to an inertial system.

\subsection{Dispersionless case and metric}

Assume, for simplicity, a dispersionless case
where the refractive index $n_0$  of the probe
does not depend on the frequency.
Note that a horizon inevitably violates this condition,
because here light comes to a standstill, oscillating at
increasingly shorter wavelenghts
the closer it approaches the horizon.
Light waves are dramatically frequency shifted
and thus leave any dispersionless frequency window.
However, many of the essentials of horizons are still
captured within the simplified dispersionless model.

First, we can cast the wave equation
(\ref{eq:wave}) in a relativistic form,
introducing a relativistic notation
\cite{LL2} for the coordinates and their derivatives
 %%%%%%%%%%%%%%%%%%
\begin{equation}
x^\mu = (\zeta,\tau) \,,\quad
\partial_\mu = (\partial_\zeta, \partial_\tau)
\end{equation}
%%%%%%%%%%%%%%%%%%
and the matrix
 %%%%%%%%%%%%%%%%%%
\begin{equation}
\label{eq:gmunu}
 g^{\mu \nu} = \left(
    \begin{array}{cc}
      1  & -1 \\
      -1 & 1-u^2n^2/c^2
    \end{array}
\right)
\end{equation}
%%%%%%%%%%%%%%%%%%
that resembles the inverse metric tensor
of waves in moving fluids \cite{UV}.
Adopting these definitions and Einstein's
summation convention over repeated indices
the wave equation (\ref{eq:wave}) appears as
 %%%%%%%%%%%%%%%%%%
\begin{equation}
\label{eq:wave1}
\partial_\mu g^{\mu\nu}\partial_\nu A = 0 \,,
\end{equation}
%%%%%%%%%%%%%%%%%%
which is almost the free wave equation in a
curved space-time geometry \cite{LL2}.
(In the case of a constant refractive index
the analogy between the moving medium
and a space-time manifold is perfect
\cite{Remark_g}.)
The effective metric tensor $g_{\mu\nu}$
is the inverse of $g^{\mu\nu}$ \cite{LL2}.
We obtain
 %%%%%%%%%%%%%%%%%%
\begin{equation}
\label{eq:metric}
g_{\mu \nu} =
\frac{c^2}{n^2u^2}
\left(
    \begin{array}{cc}
      u^2n^2/c^2-1  & -1 \\
      -1 & -1
    \end{array}
\right)
\,.
\end{equation}
%%%%%%%%%%%%%%%%%%
In subluminal regions where the velocity $c/n$ of
the probe light exceeds the speed of the
effective medium,
{\it i.e.} the velocity $u$ of the pulse,
the measure of time $u^2n^2/c^2-1$
in the metric (\ref{eq:metric}) is negative.
Here both $\partial_\tau$ and $\partial_\zeta$
are timelike vectors \cite{LL2}.
In superluminal regions, however,
$c/n$ is reduced such that $u^2n^2/c^2-1$ is positive.
A horizon, where time stands still, is established
where the velocity of light matches the speed of the pulse.

\subsection{Action}

The theory of quantum fields at horizons
\cite{Hawking,BirrellBrout}
predicts the spontaneous generation  of particles.
In our case, the quantum field is light in dielectric media.
The quantum theory of light in media at rest
has reached a significant level of sophistication \cite{Media},
because it forms the foundation of
quantum optics and, in particular, the quantum theory
of optical instruments \cite{LeoReview,Thesis},
but quantum light in moving media
is much less studied \cite{LeoReview}.
In optical fibers, light is subject to dispersion,
which represents experimental opportunities on one side,
but poses a theoretical challenge on the other:
we should quantize a field described by a
classical wave equation of high order in the retarded time.
Moreover, strictly speaking,
dispersion is always accompanied by dissipation,
which results in additional quantum fluctuations
\cite{Media}.
Here, however, we assume to operate in frequency windows
where the absorption is very small.
We entirely focus on the dispersive properties of the fibre.
To deduce the starting point of the theory,
we begin with the dispersionless case in classical optics
and then proceed to consider
optical dispersion for light quanta.

The classical wave equation of
one-dimensional light propagation in dispersionless media
follows from the Principle of Least Action \cite{LL2} with
the action of the electromagnetic field in SI units
%%%%%%%%%%%%%%%%%%
\begin{eqnarray}
S &=& \int\int \frac{\varepsilon_0}{2}
\big(n^2E^2 - c^2 B^2\big) dz dt
\nonumber\\
&=& \int\int \frac{\varepsilon_0}{2}
\big[n^2(\partial_t A)^2 - c^2 (\partial_z A)^2\big] dz dt
\nonumber\\
&=& \int\int \frac{\varepsilon_0}{2}
\big[-A\partial_\tau n^2\partial_\tau A -
c^2 (\partial_z A)^2\big] u\, d\tau d\zeta
\end{eqnarray}
%%%%%%%%%%%%%%%%%%
and hence the Lagrangian density
%%%%%%%%%%%%%%%%%%
\begin{equation}
{\mathscr L} = -\frac{\varepsilon_0}{2}
\big[A\partial_\tau n^2\partial_\tau A +
c^2 (\partial_z A)^2\big]
\,.
\end{equation}
%%%%%%%%%%%%%%%%%%
In order to include the optical dispersion
in the fiber and the effect of the moving pulse,
we express the refractive index in terms of
$\beta(\omega)$ and the effective susceptibility $\chi(\tau)$
caused by the pulse,
using Eqs.\ (\ref{eq:beta}) and (\ref{eq:n}) with
$\omega=i\partial_\tau$.
We thus propose the Lagrangian density
%%%%%%%%%%%%%%%%%%
\begin{equation}
\label{eq:lagrangian}
{\mathscr L} = \frac{\varepsilon_0}{2}
\big[A\big(c^2\beta^2(i\partial_\tau)
-\partial_\tau \chi \partial_\tau \big)A -
c^2 (\partial_z A)^2\big]
\,.
\end{equation}
%%%%%%%%%%%%%%%%%%
In the absence of losses, $\beta^2(\omega)$ is an even function,
as we obtained in Sec.\ A.2.
We write down
the Euler-Lagrange equation \cite{LL2}
for this case
%%%%%%%%%%%%%%%%%%
\begin{equation}
\partial_\zeta
\frac{\partial{\mathscr L}}{\partial(\partial_\zeta A)} -
\sum_{\nu=0}^\infty (-1)^\nu \partial_\tau^\nu
\frac{\partial{\mathscr L}}{\partial(\partial_\tau^\nu A)}
= 0
\end{equation}
%%%%%%%%%%%%%%%%%%
and obtain the wave equation (\ref{eq:wave}).
This proves that the Lagrangian density
(\ref{eq:lagrangian}) is the correct one.

\subsection{Quantum field theory}

According to the quantum theory of fields \cite{Weinberg}
the component $A$ of the vector potential is
described by an operator $\hat{A}$.
Since the classical field $A$ is real, the operator $\hat{A}$
must be Hermitian.
For finding the dynamics of the quantum field
we quantize the classical relationship
between the field, the canonical momentum density
and the Hamiltonian:
we replace the Poisson bracket between
the field $A$ and the momentum density
$\partial{\mathscr L}/\partial(\partial_\zeta A)$
by the fundamental commutator between
the quantum field $\hat{A}$
and the quantized momentum density \cite{Weinberg}.
We obtain from the Lagrangian (\ref{eq:lagrangian})
the canonical momentum density
%%%%%%%%%%%%%%%%%%
\begin{equation}
\hat{\pi} = -\varepsilon_0 \frac{c^2}{u}\,\partial_z \hat{A}
\end{equation}
%%%%%%%%%%%%%%%%%%
and postulate the equivalent of the
standard equal-time commutation relation
\cite{Weinberg,MandelWolf}
%%%%%%%%%%%%%%%%%%
\begin{equation}
\label{eq:comm}
\big[\hat{A}(\zeta,\tau_1),\hat{\pi}(\zeta,\tau_2)\big] =
\frac{i\hbar}{u}\,\delta(\tau_1-\tau_2) \,.
\end{equation}
%%%%%%%%%%%%%%%%%%
We obtain the Hamiltonian
%%%%%%%%%%%%%%%%%%
\begin{eqnarray}
\hat{H} &=& \int \big(
\hat{\pi}\partial_\zeta\hat{A} - \hat{\mathscr L}
\big) u\, d\tau
\nonumber\\
&=&
\frac{\varepsilon_0}{2} \int
\bigg(\frac{c^2}{u^2} \Big((\partial_\tau \hat{A})^2 -
(\partial_\zeta \hat{A})^2 \Big)
-\hat{A}\Big(c^2\beta^2-\partial_\tau \chi \partial_\tau\Big)\hat{A}
\bigg)
u\, d\tau \,.
\label{eq:hamilton}
\end{eqnarray}
%%%%%%%%%%%%%%%%%%
One verifies that the Heisenberg equation of the quantum
field $\hat{A}$ is the classical wave equation (\ref{eq:wave}),
as we would expect for fields that obey linear field equations.

\subsection{Mode expansion}

Since the field equation is linear and classical,
we represent $\hat{A}$
as a superposition of a complete set of classical modes
multiplied by quantum amplitudes $\hat{a}_k$.
The mode expansion is Hermitian
for a real field such as the electromagnetic field,
 %%%%%%%%%%%%%%%%%%
\begin{equation}
\label{eq:modes}
\hat{A} = \sum_k\left(
A_k\hat{a}_k + A^*_k \hat{a}_k^\dagger\right) \,.
\end{equation}
%%%%%%%%%%%%%%%%%%
The modes $A_k$ obey the classical wave equation
(\ref{eq:wave1})
and are subject to the orthonormality relations
\cite{BirrellBrout,LeoReview}
 %%%%%%%%%%%%%%%%%%
\begin{equation}
\label{eq:on}
\left(A_k,A_{k'}\right)=\delta_{kk'} \,\quad
\left(A_k^*,A_{k'}\right)=0
\end{equation}
%%%%%%%%%%%%%%%%%%
with respect to the scalar product
%%%%%%%%%%%%%%%%%%
\begin{equation}
\label{eq:scalar}
\left(A_1,A_2\right)=
\frac{\varepsilon_0 c^2}{i\hbar}\int\left(
A_1^*\partial_z A_2 - A_2\partial_z A_1^*\right) d\tau
\,.
\end{equation}
%%%%%%%%%%%%%%%%%%
The scalar product is chosen such that it is a conserved quantity
for any two solutions $A_1$ and $A_2$
of the classical wave equation (\ref{eq:wave}),
%%%%%%%%%%%%%%%%%%
\begin{equation}
\partial_\zeta(A_1,A_2)=0
\,,
\end{equation}
%%%%%%%%%%%%%%%%%%
with a prefactor that
turns out to make the commutation relations between the mode
operators particularly simple and transparent.

The scalar product serves to identify the quantum
amplitudes $\hat{a}_k$ and $\hat{a}_k^\dagger$:
the amplitude $\hat{a}_k$ belongs to modes $A_k$
with positive norm, whereas the Hermitian conjugate
$\hat{a}_k^\dagger$ is the quantum amplitude to
modes $A_k^*$ with negative norm,
because
 %%%%%%%%%%%%%%%%%%
\begin{equation}
\left(A_1^*,A_2^*\right)=-\left(A_1,A_2\right)\,.
\end{equation}
%%%%%%%%%%%%%%%%%%
Using the orthonormality relations (\ref{eq:on})
we can express the mode operators $\hat{a}_k$
and $\hat{a}_k^\dagger$ as projections of the quantum
field $\hat{A}$ onto the modes $A_k$ and $A_k^*$
with respect to the scalar product (\ref{eq:scalar}),
%%%%%%%%%%%%%%%%%%
\begin{equation}
\hat{a}_k = \big(A_k,\hat{A}\big)
\,,\quad
\hat{a}_k^\dagger = -\big(A_k^*,\hat{A}\big)
\,.
\end{equation}
%%%%%%%%%%%%%%%%%%
We obtain from the fundamental commutator (\ref{eq:comm})
and the orthonormality relations (\ref{eq:on}) of the modes
the Bose commutation relations
%%%%%%%%%%%%%%%%%%
\begin{equation}
[\hat{a}_k,\hat{a}_{k'}^\dagger] = \delta_{kk'} \,,\quad
[\hat{a}_k,\hat{a}_{k'}] = 0 \,,
\end{equation}
%%%%%%%%%%%%%%%%%%
which justifies the choice of the prefactor
in the scalar product (\ref{eq:scalar}).
In agreement with the spin-statistics theorem
\cite{Weinberg}, light consists of bosons
and the quantum amplitudes $\hat{a}_k$
and $\hat{a}_k^\dagger$
serve as annihilation and creation operators.

The expansion (\ref{eq:modes}) is valid for any
orthonormal and complete set of modes.
Consider stationary modes with frequencies $\omega_k'$
such that
%%%%%%%%%%%%%%%%%%
\begin{equation}
\partial_\zeta A_k = -i\omega_k' A_k \,.
\end{equation}
%%%%%%%%%%%%%%%%%%
We substitute the mode expansion (\ref{eq:modes})
in the Hamiltonian (\ref{eq:hamilton}) and use the
wave equation (\ref{eq:wave}) and the
orthonormality relations (\ref{eq:on}) to obtain
%%%%%%%%%%%%%%%%%%
\begin{equation}
\hat{H} = \hbar \sum_k \omega_k' \left(
\hat{a}_k^\dagger\hat{a}_k + \frac{1}{2}
\right) \,.
\end{equation}
%%%%%%%%%%%%%%%%%%
Each stationary mode contributes $\hbar\omega_k'$
to the total energy that also includes the vacuum energy.

The modes with positive norm select the
annihilation operators of a quantum field,
whereas the negative norm modes pick out
the creation operators.
In other words, the norm of the modes determines
the particle aspects of the quantum field.
In the Unruh effect \cite{Uff},
modes with positive norm in the Minkowski
space-time consist of superpositions of
positive and negative norm modes in the frame
of an accelerated observer \cite{BirrellBrout}.
Consequently, the Minkowski vacuum is not the
vacuum as seen in the accelerated frame.
Instead, the accelerated observer perceives
the Minkowski vacuum as thermal radiation \cite{Uff}.
In the Hawking effect \cite{Hawking},
the scattering of light at the event horizon
turns out to mix positive and negative norm modes,
giving rise to Hawking radiation.

\subsection{Geometrical optics}

A moving dielectric medium with constant refractive index
but nonuniform velocity appears to light exactly
as an effective space-time geometry
\cite{LeoReview,Remark_g}.
Since a stationary $1+1$ dimensional geometry
is conformally flat \cite{Curvature}
a coordinate transformation can reduce the
wave equation to describing wave propagation
in a uniform medium, leading to plane-wave solutions
\cite{GREE}.
The plane waves appear as phase-modulated waves
in the original frame.
Consequently, in this case, geometrical optics is exact.
In our case, geometrical optics provides an
excellent approximation,
because the variations of the refractive index are
very small.

Consider a stationary mode $A$.
We assume that the mode carries a slowly varying
amplitude ${\cal A}$ and oscillates with a rapidly
changing phase $\varphi$,
 %%%%%%%%%%%%%%%%%%
\begin{equation}
\label{eq:go}
A = {\cal A} \exp(i\varphi) \,.
\end{equation}
%%%%%%%%%%%%%%%%%%
We represent the phase as
 %%%%%%%%%%%%%%%%%%
\begin{equation}
\label{eq:phase}
\varphi = - \int \omega(\tau)\, d\tau - \omega' \zeta
\end{equation}
%%%%%%%%%%%%%%%%%%
and obtain from the wave equation (\ref{eq:wave})
the dispersion relation
%%%%%%%%%%%%%%%%%%
\begin{equation}
\label{eq:dispersion}
(\omega-\omega')^2 = \frac{u^2}{c^2}\,n^2\omega^2
\end{equation}
%%%%%%%%%%%%%%%%%%
by neglecting all derivatives of the amplitude ${\cal A}$.
Here $n$ includes the additional susceptibility $\chi$
due to the Kerr effect of the pulse according to Eq. (\ref{eq:n}).

The dispersion relation has two sets of solutions describing
waves that are co- or counter-propagating
with the pulse in the laboratory frame.
Counter-propagating waves will experience the pulse as
a tiny transient change of the refractive index,
whereas co-propagating modes may be profoundly affected.
Consider the solution
%%%%%%%%%%%%%%%%%%
\begin{equation}
\label{eq:doppler}
\omega'= \left(1-\frac{u}{c}n\right)\omega \,.
\end{equation}
%%%%%%%%%%%%%%%%%%
In this case, we obtain
outside of the pulse in the laboratory frame
$\varphi = n(\omega/c) z - \omega t$,
which describes light propagating in the positive $z$ direction.
Consequently, the branch (\ref{eq:doppler})
of the dispersion relation
corresponds to co-propagating light waves.
We also see that $\omega$
is the frequency of light in the laboratory frame,
whereas $\omega'$ is the frequency in the
frame co-moving with the pulse.
Equation (\ref{eq:doppler}) describes how
the laboratory-frame and the co-moving frequencies
are connected due to the Doppler effect.

In order to find the evolution of the amplitude ${\cal A}$,
we substitute in the exact scalar product  (\ref{eq:scalar})
the approximation (\ref{eq:go})
with the phase (\ref{eq:phase})
and the dispersion relation (\ref{eq:doppler}).
In the limit $\omega_1'\rightarrow\omega_2'$
we obtain
%%%%%%%%%%%%%%%%%%
\begin{equation}
\left(A_1,A_2\right)=
\frac{2\varepsilon_0 c}{\hbar}\int
{\cal A}^2 n\omega\,\exp\big(i\varphi_2-i\varphi_1\big) d\tau
\,,
\end{equation}
%%%%%%%%%%%%%%%%%%
which should give $\delta(\omega_1'-\omega_2')$
according to the normalization (\ref{eq:on}).
The dominant, diverging contribution to this integral,
generating the peak of the delta function, stems
from $\tau\rightarrow\pm\infty$ \cite{LL3}.
Hence, for $\omega_1'\rightarrow\omega_2'$,
we replace $\varphi$ in the integral by $\varphi$ at
$\tau\rightarrow\pm\infty$ where $\omega$
does not depend on $\tau$ anymore,
%%%%%%%%%%%%%%%%%%
\begin{equation}
\left(A_1,A_2\right)=
\frac{2\varepsilon_0 c}{\hbar}\int
{\cal A}^2 n\omega\,\exp\big[i(\omega_2-\omega_1)\tau\big] d\tau
\,,\quad
\omega_2-\omega_1
= \frac{\partial \omega}{\partial \omega'}(\omega_2'-\omega_1') \,,
\end{equation}
%%%%%%%%%%%%%%%%%%
which gives $\delta(\omega_1'-\omega_2')$ for
%%%%%%%%%%%%%%%%%%
\begin{equation}
\label{eq:amplitude}
|{\cal A}|^2 = \frac{\hbar}{4\pi\varepsilon_0 c n \omega}\,
\left|\frac{\partial\omega}{\partial\omega'}\right|
\end{equation}
%%%%%%%%%%%%%%%%%%
and positive frequencies $\omega$ in the laboratory frame.
Note that positive frequencies $\omega'$
in the co-moving frame
correspond to negative $\omega$
in superluminal regions where the pulse moves faster
than the phase-velocity of the probe light.

Hamilton's equations \cite{LL1} determine
the trajectories of light rays in the co-moving frame,
parameterized by the pulse-propagation time $\zeta$.
Here $\tau$ plays the role of the ray's position.
Comparing the phase (\ref{eq:phase}) with
the standard structure of the eikonal in geometrical optics
\cite{BornWolf} or the semiclassical wave function
in quantum mechanics \cite{LL3}
we notice that $-\omega$
plays the role of the conjugate momentum here.
Therefore, we obtain Hamilton's equations
with a different sign than usual \cite{LL1},
%%%%%%%%%%%%%%%%%%
\begin{equation}
\label{eq:he}
\dot{\tau} = -\frac{\partial\omega'}{\partial\omega}
\,,\quad
\dot{\omega} = \frac{\partial\omega'}{\partial \tau}
\,.
\end{equation}
%%%%%%%%%%%%%%%%%%
We express $\dot{\tau}$ in terms of the group index
in the laboratory frame.
Here the group velocity $v_g$ is the derivative of the
frequency $\omega$ with respect to the wave number
$n\omega/c$ or, equivalently, the inverse of the
derivative of $n\omega/c$ with respect to $\omega$,
which gives for the group index $c/v_g$
the standard expression \cite{Agrawal}
%%%%%%%%%%%%%%%%%%
\begin{equation}
n_g = n + \omega\frac{\partial n}{\partial\omega}
\,.
\end{equation}
%%%%%%%%%%%%%%%%%%
We obtain from the first of Hamilton's equations (\ref{eq:he})
and the Doppler formula (\ref{eq:doppler})
%%%%%%%%%%%%%%%%%%
\begin{equation}
\label{eq:vg}
\dot{\tau} = \frac{u}{c}\,n_g-1
= -\frac{n_g}{c} v_g'
\,,\quad v_g' = \frac{c}{n_g}-u
\end{equation}
%%%%%%%%%%%%%%%%%%
where $v_g'$ denotes the difference between
the group velocity of the probe $v_g$
and the pulse speed $u$.
We see that the velocity $\dot{\tau}$ in the co-moving frame
(\ref{eq:frame}) vanishes when the Kerr susceptibility $\chi$
reduces the group velocity $c/n_g$
such that it matches the speed of the pulse $u$.
Since $\dot{\omega}$ does not vanish here in general,
the ray does not remain there,
but changes direction in the co-moving frame.

At such a turning point we expect a violation of the validity
of geometrical optics \cite{LL3}.
For example, the amplitude (\ref{eq:amplitude})
would diverge here.
Geometrical optics is an exponentially accurate approximation
when
%%%%%%%%%%%%%%%%%%
\begin{equation}
\left|\frac{\partial T}{\partial \tau}\right| \ll 1
\quad\mbox{for}\quad
T = \frac{2\pi}{\omega}
\,,
\end{equation}
%%%%%%%%%%%%%%%%%%
as we see from the analogy to the semiclassical approximation
in quantum mechanics \cite{LL3}.
Here the cycle $T$ plays the role of the wavelength.
We get
%%%%%%%%%%%%%%%%%%
\begin{equation}
\frac{\partial T}{\partial \tau}= \frac{\dot{\omega}T}{\omega\dot{\tau}}
\,.
\end{equation}
%%%%%%%%%%%%%%%%%%
Consequently, geometrical optics indeed is no longer valid near
a turning point where
%%%%%%%%%%%%%%%%%%
\begin{equation}
n_g = \frac{c}{u}
\,.
\end{equation}
%%%%%%%%%%%%%%%%%%
This turning point defines a {\it group-velocity horizon}
where the pulse has slowed down the probe such
that it matches the speed of the pulse.
At this horizon the incident mode is converted into a mode
that represents another solution of the dispersion relation;
a red-or blue-shifted wave, depending on the dispersion
and the sign of the first derivative of $\chi$ with respect to $\tau$
at the group-velocity horizon.
White holes correspond to increasing $\chi$ and black holes
to decreasing $\chi$.
White holes blue-shift, because incident waves freeze in front
of the horizon, oscillating with increasing frequency.
Black holes red-shift, because they stretch any emerging waves
(also because black holes are time-reversed white holes).
Due to the effective dispersion of the fiber,
the refractive index changes with frequency.
In turn, the dispersion limits the frequency shifting by
tuning the light out of the grip of the horizon.
In particular, the dispersion limits the blue-shifting
at white-hole horizons to respectable but finite frequencies,
considering the tiny magnitude of $\chi$,
as we discuss in Sec.\ B2.

At the event horizons of astrophysical black holes,
similar effects are expected \cite{TransPlanck}
when, due to the wave-number divergence, the
wavelength of light is reduced below the
Planck length scale $\sqrt{h G /c^3}$
where $G$ is the gravitational constant.
The physics beyond the Planck scale is unknown.
This trans-Planckian physics
should regularize the logarithmic phase singularities
\cite{BirrellBrout}
of modes at the event horizon.
A numerical study of a simple model of
trans-Planckian-type physics \cite{UnruhTP}
and a systematic analytical study \cite{Brout2}
indicate that the Hawking effect of the black hole
is not affected.
On the other hand, the quantum radiation
of white holes is dominated by
trans-Planckian physics,
because of the extreme blue shift
at white-hole horizons.
It has been predicted \cite{CJ} that
black-hole white-hole pairs could act
as black hole lasers in a regime
of anomalous group-velocity dispersion.
From a theoretical point of view,
trans-Planckian physics
regularizes some of the arcane features
of quantum black holes and gives a more
natural picture of the physics behind the Hawking effect
\cite{Brout2}.
In our case,
the optical analogue of trans-Planckian physics,
optical dispersion, is known in principle
and turns out to be to the advantage of the experiment.

\subsection{Classical Hawking effect}

A {\it phase-velocity horizon} is formed
if the pulse has slowed down the probe
such that its phase velocity is lower than the speed of the pulse.
Here an additional effect occurs: the spontaneous creation of
photon pairs, Hawking radiation.

\begin{figure}[t]
\begin{center}
\includegraphics[width=20.0pc]{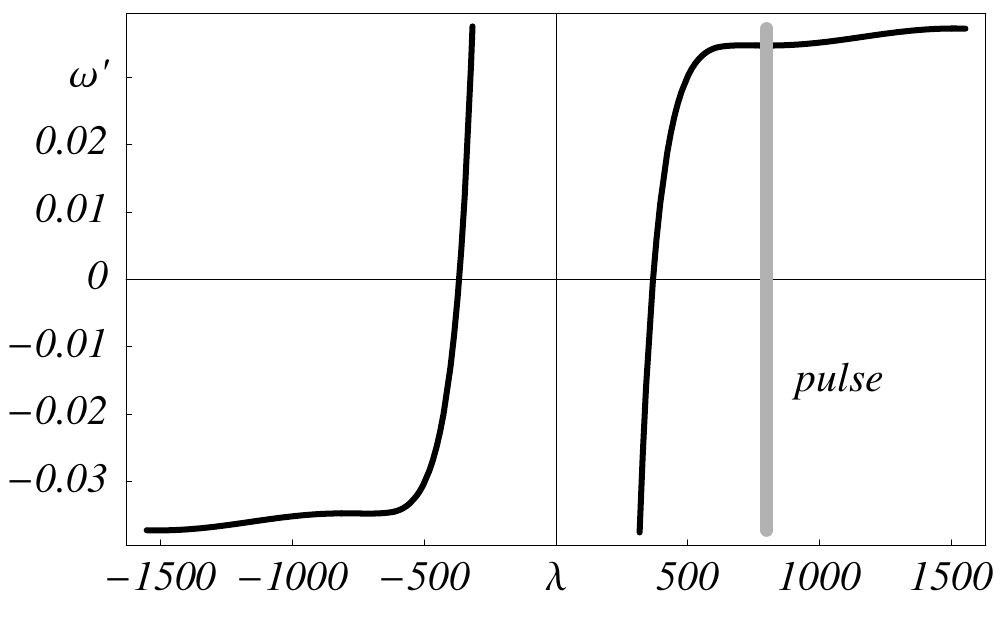}
\caption{
\small{
Doppler-shifted frequency versus wavelength.
The figure shows the co-moving frequencies (\ref{eq:doppler})
in units of $10^{15}\mathrm{Hz}$
for our micro-structured fiber
depending on the wavelengths $\lambda = 2\pi/\omega$
in $\mathrm{nm}$.
To determine the effective refractive index we used
the dispersion data of our fiber in
Eq.\ (\ref{eq:betadispersion}), apart from two constants
that we fitted to the Sellmeier formula
for fused silica at short wavelengths \cite{Agrawal}.
}
\label{fig:map}}
\end{center}
\end{figure}

In the near ultraviolet around $300\mathrm{nm}$,
the dispersion of microstructured fibers  \cite{Russells}
is dominated by the bare dispersion of glass where
$n_0$ rapidly grows with frequency \cite{Agrawal},
exceeding the group index $c/u$ of the pulse.
For such ultraviolet modes,
the medium moves at superluminal speed.
According to the Doppler formula (\ref{eq:doppler})
these superluminal modes oscillate with negative
frequencies $\omega'$ in the co-moving frame
for positive frequencies $\omega$ in the laboratory frame,
and vice versa.
On the other hand, probe modes with phase velocities
below $c/n$ oscillate with positive frequencies.
Therefore,
two waves share a given  $\omega'$ in the co-moving frame,
a subluminal wave with positive frequency $\omega$
in the laboratory frame
and a superluminal wave with negative $\omega$,
see Fig.\ \ref{fig:map}.
In the case of the astrophysical event horizon,
the positive-frequency modes correspond to waves outside
the horizon that escape into space,
and the negative-frequency modes to waves
beyond the horizon that fall into the singularity \cite{BirrellBrout}.
The Kerr susceptibility of the pulse may slow down
the subluminal modes such that the pulse moves at superluminal speed.
As we will show,
in this case sub- and superluminal modes
are partially converted into each other.
This mode conversion is at the heart of the Hawking effect.

%%%
%%%

For simplicity, we consider a single white-hole horizon,
not the combination of black- and white-hole horizons generated
by a moving pulse.
We will argue later that in practice
the white-hole will dominate the Hawking effect,
which {\it a-posteriori} justifies this simplification.
Suppose,
without loss of generality, that at $\tau=0$
the Kerr-reduced phase-velocity of the probe, $c/n$,
matches the group velocity of the pulse $u$.
We assume that the mode conversion occurs
near this point
and expand the Kerr susceptibility $\chi$ as a linear function
in $\tau$,
%%%%%%%%%%%%%%%%%%
\begin{equation}
\label{eq:linear}
\chi(\tau) = \chi_h + \alpha''\tau
\,,\quad
\alpha'' = \left.\frac{\partial\chi}{\partial\tau}\right|_0
\,.
\end{equation}
%%%%%%%%%%%%%%%%%%
The group velocity of the incident probe
is much lower than the pulse speed $u$
and so both the sub- and the superluminal probe
travels from the front of the pulse to the back,
from negative to positive retarded time $\tau$.
For a white-hole horizon $\chi$ increases
for decreasing retarded time, and so $\dot{\chi}(0)<0$.

We proceed similar to Ref.\ \cite{Brout2} and
focus on the conversion region
where we Fourier-transform with respect to $\tau$
the wave equation (\ref{eq:wave})
with the refractive index (\ref{eq:n})
for stationary waves in the co-moving frame
and using
the linear expansion (\ref{eq:linear}).
The frequency conjugate to $\tau$ is the laboratory-frame
frequency $\omega$.
We replace $\tau$ by $-i\partial_\omega$,
$\partial_\zeta$ by $-i\omega'$ and
$\partial_\tau$ by $-i\omega$,
denote the Fourier-transformed vector potential by
$\widetilde{A}$, and obtain
%%%%%%%%%%%%%%%%%%
\begin{equation}
\left(n_0^2 + \chi_h - i\alpha''\partial_\omega\right)
\omega\widetilde{A}
= \left(1-\frac{\omega'}{\omega}\right)^2 \frac{c^2}{u^2}\,
\omega\widetilde{A}
\,.
\end{equation}
%%%%%%%%%%%%%%%%%%
This first-order equation has the exact solution
%%%%%%%%%%%%%%%%%%
\begin{equation}
\widetilde{A} = \frac{\widetilde{\cal A}_0}{\omega}\, e^{-i\phi}
\,,\quad
\phi = -\frac{1}{\alpha''}
\int\left[\left(1-\frac{\omega'}{\omega}\right)^2 \frac{c^2}{u^2}
-n_0^2-\chi_h\right] d \omega
\end{equation}
%%%%%%%%%%%%%%%%%%
with constant $\widetilde{\cal A}_0$.
We introduce
%%%%%%%%%%%%%%%%%%
\begin{equation}
\alpha' \equiv - \frac{u^2}{c^2}\, \frac{\alpha''}{2}
\approx -\frac{u}{c}\, \frac{\dot{\chi}(0)}{2n_0}
= \left.-\frac{u}{c}\, \frac{\partial n}{\partial\tau}\right|_0
\end{equation}
%%%%%%%%%%%%%%%%%%
in agreement with Eq.\ (4) of our paper \cite{Philbin}.
Note that the phase $\phi$ contains a logarithmic contribution,
%%%%%%%%%%%%%%%%%%
\begin{equation}
\label{eq:log}
\phi = -\frac{\omega'}{\alpha'}\,\ln\omega + \phi_0(\omega)
\,.
\end{equation}
%%%%%%%%%%%%%%%%%%
This logarithmic asymptotics of the phase will lead to the
characteristic mode conversion at the group-velocity horizon.
In order to see this, we Fourier-transform $\widetilde{A}$
back to the domain of the retarded time,
%%%%%%%%%%%%%%%%%%
\begin{equation}
A = \int_{-\infty}^{+\infty} \widetilde{A}\,
e^{-i\omega\tau} d\omega
= \int_{-\infty}^{+\infty}
\frac{\widetilde{\cal A}_0}{\omega}\, e^{-i\phi-i\omega\tau}
d\omega
\end{equation}
%%%%%%%%%%%%%%%%%%
and use the saddle-point approximation,
{\it i.e.}\ we quadratically expand the phase $\phi+\omega\tau$
around the stationary points where $\partial_\omega(\phi+\omega\tau)$ vanishes and perform the integration as Gaussian integrals
along the direction of steepest descent.
One easily verifies that the stationary points
are the solutions of the dispersion relation (\ref{eq:dispersion}).
We denote the two solutions by $\omega_\pm$
indicating their sign.
We obtain for the second derivative in the quadratic expansion
%%%%%%%%%%%%%%%%%%
\begin{equation}
\partial_\omega^2(\phi+\omega\tau)
= -\frac{2}{\dot{\chi}(0)}\,\frac{n}{\omega}\left(\frac{c}{u}-n_g\right)
\,.
\end{equation}
%%%%%%%%%%%%%%%%%%
The Gaussian integrals at $\omega_\pm$ are
proportional to the inverse square root of
$\partial_\omega^2(\phi+\omega\tau)$.
We see from Eqs.\ (\ref{eq:he}) and (\ref{eq:vg})
that they are consistent with the amplitudes
(\ref{eq:amplitude}) of geometrical optics.
Consequently, we obtain a superposition
of the two waves (\ref{eq:go}) that correspond to
the two physically-relevant
branches of the dispersion relation (\ref{eq:dispersion}).
We denote the positive-frequency wave by $A_+$ and the
negative-frequency component by $A_-^*$.
The star indicates that this component resembles
the complex conjugate of a mode,
because a mode predominantly contains positive
laboratory-frame frequencies, according to the normalization
(\ref{eq:amplitude}).
The coefficient of $A_-^*$
is given by the exponential of the phase integral from
the positive branch $\omega_+$ to the negative frequency
$\omega_-$ on the complex plane.
The amplitude of the coefficient is the exponent
of the imaginary part of the phase integral,
while the phase of the coefficient is given by the real part.
We can incorporate the phase of the superposition coefficient
in the prefactor (\ref{eq:amplitude}),
but not the amplitude.
The imaginary part of the phase integral comes from the
logarithmic term (\ref{eq:log}), giving $\pi\omega'/\alpha'$.
Therefore, the relative weight of the negative-frequency component
in the converted mode
is $\exp(-\pi\omega'/\alpha')$.
We thus obtain for $\tau<0$
%%%%%%%%%%%%%%%%%%
\begin{equation}
A \sim Z^{1/2}\left(A_+ + A_-^* e^{-\pi\omega'/\alpha'}\right)
\end{equation}
%%%%%%%%%%%%%%%%%%
where $Z$ denotes a constant for given $\omega'$.
We determine the physical meaning of $Z$ in Sec.\ A10,
but here we can already work out its value
by the following procedure:
consider a wavepacket with co-moving frequencies around $\omega'$
that crosses the horizon.
Suppose that this wavepacket is normalized to unity.
After having crossed the horizon,
the norm of the positive-frequency component is $Z$,
while the negative-frequency component has the negative norm
$-Z\exp(-2\pi\omega'/\alpha')$.
The sum of the two components must give unity, and so
%%%%%%%%%%%%%%%%%%
\begin{equation}
\label{eq:z}
Z = \left(1 - e^{-2\pi\omega'/\alpha'}\right)^{-1}
\,.
\end{equation}
%%%%%%%%%%%%%%%%%%
We represent $Z^{1/2}$ as $ \cosh\xi$
and obtain from Eq.\ (\ref{eq:z}) that $\sinh^2\xi = Z-1$
gives $Z e^{-2\pi\omega'/\alpha'}$,
%%%%%%%%%%%%%%%%%%
\begin{equation}
Z^{1/2} = \cosh\xi \,,\quad
Z^{1/2}e^{-\pi\omega'/\alpha'} = \sinh\xi
\,.
\end{equation}
%%%%%%%%%%%%%%%%%%
Consequently, the incident wave $A_\pm$
is converted into the superposition
$A_\pm \cosh\xi + A^*_\mp\sinh\xi$
when it crosses the horizon
from positive to negative $\tau$.
Hence we obtain for this process the mode %%%%%%%%%%%%%%%%%%
\begin{equation}
\label{eq:in}
A_{\pm\mathrm{in}}
\sim \left\{
    \begin{array}{lcr}
    A_\pm &:& \tau>0 \\
    A_\pm \cosh\xi + A^*_\mp\sinh\xi &:& \tau<0
    \end{array}
\right.
\,.
\end{equation}
%%%%%%%%%%%%%%%%%%
Equation (\ref{eq:in}) describes the fate of a classical wave
that crosses the horizon.
A negative-frequency component is generated with weight $\sinh^2\xi$
relative to the initial wave,
but, since $\cosh\xi>1$, the positive-frequency wave has been amplified.
The mode conversion at the horizon is thus an unusual scattering process
where the concerted modes are amplified,
at the expense of the energy of the driving mechanism,
the pulse in our case.
(It is also mathematically unusual --- the Hawking effect corresponds
to scattering without turning points in the complex plane.)
Wherever there is amplification of classical waves,
{\it i.e.} stimulated emission of waves,
there also is spontaneous emission of quanta \cite{Caves}
--- in the case of horizons,
Hawking radiation.

\subsection{Hawking radiation}

Suppose that no classical probe light is incident;
the modes $A_{\pm\mathrm{in}}$ are in the vacuum state.
The incident modes are characterized by the asymptotics
$A_\pm$ for $\tau>0$
while outgoing modes are required to approach $A_\pm$
for $\tau<0$.
We perform the superposition
%%%%%%%%%%%%%%%%%%
\begin{equation}
\label{eq:out}
A_{\pm\mathrm{out}}
= A_{\pm\mathrm{in}} \cosh\xi - A^*_{\mp\mathrm{in}}\sinh\xi
\end{equation}
%%%%%%%%%%%%%%%%%%
and see that $A_{\pm\mathrm{out}}$ obeys the asymptotics
%%%%%%%%%%%%%%%%%%
\begin{equation}
A_{\pm\mathrm{out}}
\sim \left\{
    \begin{array}{lcr}
    A_\pm \cosh\xi - A^*_\mp\sinh\xi &:& \tau>0 \\
    A_\pm &:& \tau<0
    \end{array}
\right.
\,,
\end{equation}
%%%%%%%%%%%%%%%%%%
as required for outgoing modes.
The modes (\ref{eq:in}) and (\ref{eq:out})
describe two sets of mode expansions (\ref{eq:modes})
of one and the same quantum field;
for a given $\omega'$ the sum of
$A_{\pm\mathrm{in}}\hat{a}_{\pm\mathrm{in}}$
and $A_{\pm\mathrm{in}}^*\hat{a}_{\pm\mathrm{in}}^\dagger$
over the two signs $\pm$ of $\omega$
must give the corresponding sum of $A_{\pm\mathrm{out}}\hat{a}_{\pm\mathrm{out}}$
and $A_{\pm\mathrm{out}}^*\hat{a}_{\pm\mathrm{out}}^\dagger$.
Consequently,
%%%%%%%%%%%%%%%%%%
\begin{equation}
\hat{a}_{\pm\mathrm{in}}
= \hat{a}_{\pm\mathrm{out}} \cosh\xi - \hat{a}^\dagger_{\mp\mathrm{out}}\sinh\xi
\end{equation}
%%%%%%%%%%%%%%%%%%
and by inversion
%%%%%%%%%%%%%%%%%%
\begin{equation}
\hat{a}_{\pm\mathrm{out}}
= \hat{a}_{\pm\mathrm{in}} \cosh\xi + \hat{a}^\dagger_{\mp\mathrm{in}}\sinh\xi
\,.
\end{equation}
%%%%%%%%%%%%%%%%%%
The vacuum state $|\mathrm{vac}\rangle$ of the incident field
is the eigenstate of the annihilation operators
$\hat{a}_{\pm\mathrm{in}}$ with zero eigenvalue
(the state that the $\hat{a}_{\pm\mathrm{in}}$ annihilate),
%%%%%%%%%%%%%%%%%%
\begin{equation}
\hat{a}_{\pm\mathrm{in}}
|\mathrm{vac}\rangle = 0
\,.
\end{equation}
%%%%%%%%%%%%%%%%%%
To find out whether and how many quanta
are spontaneously emitted by the horizon,
we express the in-coming vacuum
in terms of the out-going modes.
We denote the out-going photon-number eigenstates,
the out-going Fock states \cite{LeoReview},
by $|n_+,n_-\rangle$ with the integers $n_\pm$.
Using the standard relations for the annihilation and
creation operators
%%%%%%%
\begin{equation}
\hat{a}|n\rangle = \sqrt{n}\,|n-1\rangle
\,,\quad
\hat{a}^\dagger|n\rangle = \sqrt{n+1}\,|n+1\rangle
\,,
\end{equation}
%%%%%%%
one verifies that $\hat{a}_{\pm\mathrm{in}}$
vanishes for the state
%%%%%%%
\begin{equation}
\label{eq:epr}
|\mathrm{vac}\rangle = Z^{-1/2} \sum_{n=0}^\infty
e^{-n\pi\omega'/\alpha'}
|n,n\rangle \,.
\end{equation}
%%%%%%%
This is the remarkable result obtained by
Hawking \cite{Hawking} for the horizon of the black hole.
First, it shows that
the event horizon spontaneously generates radiation
from the incident quantum vacuum.
Second, the emitted radiation consists
of correlated photon pairs,
each photon on one side is correlated
to a partner photon on the other side,
because they are always produced in pairs.
The total quantum state turns out to be an
Einstein-Podolski-Rosen state \cite{LeoReview},
the strongest entangled state for
a given energy \cite{Barnett}.
Third, light on either side of the horizon
consists of an ensemble of photon-number
eigenstates with probability
$Z^{-1}e^{-2n\pi\omega'/\alpha'}$.
This is a Boltzmann distribution
of $n$ photons with energies $n\hbar\omega'$
and temperature $k_B T'= \hbar\alpha'/(2\pi)$.
Consequently, the horizon emits
a Planck spectrum of black-body radiation
with the Hawking temperature (5) of our paper \cite{Philbin}.
Fourth, this Planck spectrum is consistent
with Bekenstein's black-hole
thermodynamics \cite{Bekenstein}:
black holes seem to have an entropy and a temperature.

In our case, the spectrum of the emitted quanta
is a Planck spectrum for the frequencies $\omega'$
in the co-moving frame,
as long as a phase-velocity horizon exists.
We performed our analysis for the white-hole
horizon, but, since black holes are time-reversed white holes,
we arrive at the same result for the black hole as well,
except that the roles of the
incident and outgoing modes are reversed.
In the laboratory frame,
the spectrum is given by the dependance of
$\omega'$ on the laboratory frequency $\omega$
outside of the pulse,
{\it i.e.} by the dispersion relation (\ref{eq:doppler})
for $\chi=0$.
In our case, $\omega(\omega')$
is single-valued for the spectral region
where phase-velocity horizons are established,
see Fig.\ \ref{fig:map},
and so the spectrum of black- and white-hole
horizons is identical for identical $\alpha'$.
We have shown in Eq.\ (8)
that this mapping from $\omega$ to $\omega'$
amounts to a re-definition of the Hawking temperature:
in the laboratory frame $k_B T$ is given by
the logarithmic derivative of $\chi$ at the horizon;
it is independent of the magnitude of the Kerr susceptibility,
as long as a phase-velocity horizon is established.
The particle-production rate depends only on the
sharpness of the pulse.
This important feature makes the experimental observation
of Hawking radiation in optical fibers feasible
using few-cycle pulses \cite{Fewcycle}.

\subsection{Optical shock}

Another feature increases the Hawking temperature further:
the formation of an optical shock-front
at the trailing end of the pulse.
For a sufficiently intense pulse,
the Kerr effect strongly influences its shape;
it counteracts the dispersion of the pulse,
forming a soliton \cite{Agrawal}
and it may lead to self-steepening \cite{DeMartini}.
Consider the propagation of the pulse itself.
We describe its electric field $E$ by the envelope ${\cal E}$
and the phase at the carrier frequency $\omega_0$,
%%%%%%%
\begin{equation}
E = {\cal E} \exp\left(i\beta_0 z - i\omega_0 t\right)
\,,\quad
\beta_0 = n_0(\omega_0)\frac{\omega_0}{c}
\,,\quad
I = |{\cal E}|^2
\,.
\end{equation}
%%%%%%%
Assuming that the envelope varies over longer scales
than an optical cycle, we approximate
in the wave equation (\ref{eq:nlwave})
the differential kernel and the Kerr polarization (\ref{eq:pol}) by
%%%%%%%
\begin{eqnarray}
\left(\partial_z^2 + \beta^2 \right) E
&\approx&
\exp\left(i\beta_0 z - i\omega_0 t\right)
2\beta_0 \left(i\partial_z + \beta-\beta_0\right) {\cal E}
\,,\nonumber\\
\partial_t^2 P &\approx&
-2i\varepsilon_0\kappa\omega_0 E
\left(-i\omega_0 I + 3 \partial_t I \right)
\,.
\end{eqnarray}
%%%%%%%
The dominant nonlinear term in the
resulting wave equation is proportional to
the intensity $I$;
the term proportional to $\partial_t I$
only becomes important for sharp features
and, as we will see, leads to the formation of
an optical shock.

First we ignore the shock term and approximate
$\beta(\omega)$
around the carrier frequency $\omega_0$
by a quadratic polynomial,
%%%%%%%
\begin{equation}
\beta \approx \beta_0 + \beta_1 (\omega-\omega_0)
+ \frac{\beta_2}{2}  (\omega-\omega_0)^2
\,.
\end{equation}
%%%%%%%
Here $\beta_1$ describes the inverse of the group velocity $u$
and  $\beta_2$ the group-velocity dispersion.
Substituting $i\partial_t$ for $\omega-\omega_0$
wherever it acts on the envelope ${\cal E}$
we obtain the usual nonlinear Schr\"odinger equation
of Nonlinear Fiber Optics
\cite{Agrawal}
%%%%%%%
\begin{equation}
i\left(\partial_t + \frac{1}{\beta_1}\partial_z\right){\cal E}
- \frac{\beta_2}{2\beta_1}\partial_t^2{\cal E}
+ \frac{\omega_0^2\kappa}{c^2\beta_0\beta_1}
 |{\cal E}|^2 {\cal E} = 0
\,.
\end{equation}
%%%%%%%
This equation is integrable by the Inverse Scattering Method
\cite{Steudel}.
For anomalous group-velocity dispersion, where $\beta_2<0$,
it has the fundamental soliton solution
%%%%%%%
\begin{equation}
\label{eq:soliton}
{\cal E} =
{\cal E}_0 \,
\mathrm{sech}\left(\frac{\tau}{T_0}\right)
\exp\left(i\frac{|\beta_2|\zeta}{2\beta_1 T_0^2}\right)
\,,\quad
{\cal E}_0^2 =
\frac{c^2\beta_0 |\beta_2|}{\kappa\omega_0^2 T_0^2}
\end{equation}
%%%%%%%
in terms of the retarded time $\tau$ and the propagation time $\zeta$
according to the definition (\ref{eq:frame}).
The constant $T_0$ describes the duration of the soliton.

Consider now shock formation.
For simplicity, we ignore the group-velocity dispersion
and arrive at the nonlinear wave equation
%%%%%%%
\begin{equation}
\partial_\zeta {\cal E}
+ \gamma {\cal E}
\left(-i\omega_0 I + 3\partial_\tau I\right) = 0
\,,\quad
\gamma = \frac{\omega_0\kappa}{c^2\beta_0\beta_1}
\,.
\end{equation}
%%%%%%%
We represent ${\cal E}$ in terms of the amplitude $\sqrt{I}$
and a phase, and obtain that the evolution of the phase is
completely determined by the amplitude,
whereas the intensity obeys the transport equation
%%%%%%%
\begin{equation}
\left(\partial_\zeta + 6\gamma I \partial_\tau\right) I = 0
\,.
\end{equation}
%%%%%%%
The intensity is transported with velocity $6\gamma I$
where the retarded time $\tau$ plays the role of space
and the propagation time $\zeta$ the role of time.
In a $(\tau,\zeta)$ space-time diagram,
the initial intensity profile $I_0(\tau_0)$
is thus transported along a line
with steepness $6\gamma I(\tau_0)$.
Hence we write down the solution as
%%%%%%%
\begin{equation}
I = I_0(\tau_0)
\,,\quad
\tau - 6\gamma I_0(\tau_0)\zeta = \tau_0
\,.
\end{equation}
%%%%%%%
The relationship between $\tau$, $\zeta$ and the initial
intensity profile implicitly determines $\tau_0$.
At some $\tau$ and $\zeta$, two lines of transported intensity,
one belonging to $\tau_0$ and the other to $\tau_0'$,
may cross, developing a discontinuity, a shock.
Here
%%%%%%%
\begin{equation}
\frac{1}{\zeta} = \frac{6\gamma[I_0(\tau_0)-I_0(\tau_0')]}
{\tau_0'-\tau_0}
\rightarrow
-6\gamma\,\frac{dI_0}{d\tau_0}
\quad \mbox{for} \quad \tau_0\rightarrow\tau_0' \,.
\end{equation}
%%%%%%%
The shock time $\zeta$ is smallest
for the largest negative derivative of the initial intensity profile.
So the steepest point at the trailing end
of the initial pulse is the first to form a shock.
Although we ignored the dispersion
and Stimulated Raman Scattering \cite{Agrawal}
in this simple theory
of optical shocks \cite{DeMartini}
ultrashort pulses still form sharp features
at their trailing end,
features that may easily become comparable to the
carrier wavelength,
as we and others have seen in numerical simulations
using the method of Ref.\ \cite{Reeves}.
Therefore, the white-hole horizon will dominate
the Hawking radiation of an optical pulse
and the radiation is likely to become strong enough to be detectable.

%%%%%%%%%%%%%%%%%%%%%%%%%%%%%%%

\newpage

\renewcommand{\theequation}{B\arabic{equation}}
\setcounter{equation}{0}

\renewcommand{\thefigure}{B\arabic{figure}}
\setcounter{figure}{0}

\section{Experiment}

In this appendix we describe the experimental observation of
frequency shifting of light at the group velocity horizon. 
To our knowledge, such an experiment has never been carried 
out before; the cases closest to our scheme 
are demonstrations
of pulse trapping \cite{Nishizawa,Gorbach} 
where the dynamics are dominated by the Raman effect 
or pulse compression in a fiber grating
(optical push broom)
\cite{Pushbroom}. 
Based on the theory of Appendix A, 
we also derive mathematical expressions
for the amount of blue shifting, for the spectral shape,
and for estimating the efficiency of this process. 
We discuss the
experimental proceedings and findings and compare them with the
theory.

\subsection{Dispersion}

The creation of artificial event horizons in optical fibers
critically depends on the optical properties of these fibers.
These properties are described here and summarized in Table
\ref{tab:fiber}. To create an artificial event horizon in our
scheme, an intense optical pulse has to be formed inside the
fiber. Optical solitons \cite{Hasegawa, Agrawal} offer a unique
possibility for nondispersive stable pulses in fibers. These can
be ultrashort, allowing for very high peak powers to drive the
nonlinearity of the fiber. Bright solitons only exist for
anomalous group velocity dispersion \cite{Agrawal}.
Microstructured fibers \cite{Russells} have an arrangement of
holes close to the fiber core along the fiber. In the simplest
picture, the holes lower the local refractive index in the
transverse plane of the fiber, leading to substantially larger
index variations compared to conventional fibers. Hence a very
wide range of transverse refractive index profiles can be
engineered. There are various designs for the shape and location
of the holes, leading to a range of different effective
dispersions and giving rise to a variety of applications
\cite{Russells}. In particular, the anomalous dispersion required
for solitons can be generated at wavelengths reaching the visible.

The dispersion parameter $D$ of optical fibers is defined as the
change of group delay per wavelength change and propagation
length. Its units are usually $\mathrm{ps/(nm\,km)}$. Since the
group delay per propagation length is given by $n_g/c$ and
$n_g/c$=$\partial\beta/\partial\omega$, we have \cite{Agrawal}
%%%%%%%
\begin{equation}
\label{eq:dd} D =
\frac{\partial^2\beta}{\partial\lambda\partial\omega} \,,\quad
\lambda = \frac{2\pi c}{\omega} \,.
\end{equation}
%%%%%%%
The group-velocity dispersion is often also characterized by the
second derivative of $\beta$ with respect to the frequency
$\omega$,
%%%%%%%
\begin{equation}
\label{eq:beta2}
\beta_2 =
\frac{\partial^2\beta}{\partial\omega^2} =
-\frac{\lambda}{\omega}\, D
\,.
\end{equation}
%%%%%%%
The group-velocity dispersion is normal for positive $\beta_2$ and
negative $D$, and anomalous for negative  $\beta_2$ and positive
$D$.

\begin{figure}[t]
\begin{center}
\includegraphics[width=25.0pc]{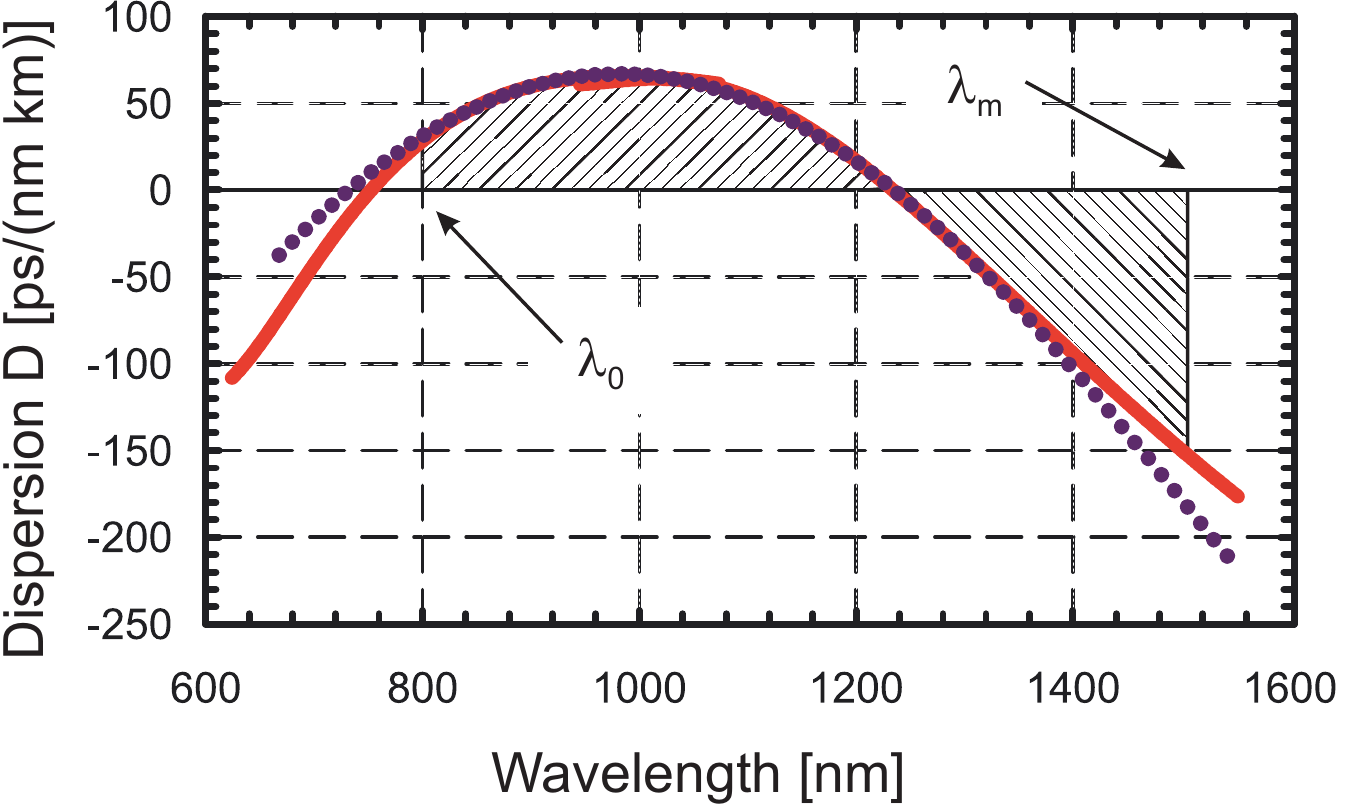}
\caption{ \small{ Two measurements of the dispersion parameter $D$
for the fiber used in the experiments. Red line: company data;
dotted line: result of Alexander Podlipensky and Philip Russell at
the Max Planck Research Group in Optics, Information and Photonics
in Erlangen, Germany. If the total shaded area vanishes, the two
wavelengths at either end are group velocity matched.}
\label{fig:dd}}
\end{center}
\end{figure}

For the creation of a horizon we chose a commercial
microstructured fiber, model NL-PM-750B by Crystal Fiber A/S.
Figure \ref{fig:dd} shows the dispersion 
of the particular fiber sample we used. The red curve is a
measurement provided by Crystal Fiber; the dotted line was
measured for our fiber sample by Alexander Podlipensky and Philip
Russell at the Max Planck Research Group in Optics, Information
and Photonics in Erlangen, Germany. The fiber dispersion is
anomalous between $\approx 740\, \mathrm{nm}$ and $\approx 1235\,
\mathrm{nm}$ wavelength and normal otherwise. Thus solitons can be
created using ultrashort pulses from modelocked Ti:Sapphire
lasers. Light that would probe the horizon and experience blue
shifting as a result, will have to be slowed down by the Kerr
effect of the pulse such that its group velocity matches the speed
$u$ of the pulse. The Kerr susceptibility is small (we give an
estimate in Sec.\ B2), and so the initial group velocity of the
probe should be only slightly higher than $u$. Integrating Eq.\
(\ref{eq:dd}) we obtain
%%%%%%%
\begin{equation}
\label{eq:gvddispersion} \int_{\lambda_0}^\lambda D\, d\lambda =
\beta_1(\lambda) - \beta_1(\lambda_0) =
\frac{1}{v_g(\lambda)}-\frac{1}{v_g(\lambda_0)}
\,,\quad v_g(\lambda_0)=u
\,.
\end{equation}
%%%%%%%
Here $\lambda_0$ and $\lambda$ denote the center wavelengths of
the pulse and the probe light, respectively. Therefore, the probe
light travels at the speed of the pulse if the integral of $D$
vanishes, as illustrated by the shaded areas in Fig.\
\ref{fig:dd}. This probe wavelength is called the 
group-velocity-matched wavelength $\lambda_m$ and the corresponding
frequency $\omega_m$ the group-velocity-matched frequency. 
For a pulse carrier-wavelength of $800\,\mathrm{nm}$ and the
fiber used here we obtain
$\lambda_m\,\!\approx\!\,1500\,\mathrm{nm}$. This value of
$\lambda_m$ is useful, because on the one hand it is a standard
wavelength for lasers and optical equipment made for
fiber-optical telecommunications and on the other
hand it is clearly separated from our broadband pulsed light.

%%%
%%%

The dispersion $D$ essentially describes the effective group index
of the fiber as a function of wavelength (or frequency).
Integrating Eq.\ (\ref{eq:dd}) twice we obtain the effective
wavenumber (\ref{eq:beta})
%%%%%%%
\begin{eqnarray}
\beta & = & \beta(\omega_0) +
\frac{\omega-\omega_0}{u}
+\int_{\omega_0}^\omega\int_{\lambda_0}^\lambda D\, d\lambda\,
d\omega = n\frac{\omega}{c}
\nonumber\\
& = & \beta(\omega_0) +
\frac{\omega-\omega_0}{u}
+\int_{\omega_0}^\omega\int_{\omega_0}^\omega 
\beta_2\, d\omega\, d\omega \,,
\label{eq:betadispersion}
\end{eqnarray}
%%%%%%%
where $n$ is the linear effective refractive index of the fiber
and $\omega_0$ and $\omega$ denote the carrier frequency of pulse
and probe, respectively. So, in addition to the measured
dispersion curve, only two constants determine $\beta$: the pulse
group velocity $u$ at $\omega_0$ and $\beta(\omega_0)$. However,
in what follows the most relevant parameters for describing our
experiment are independent of these constants.

In general, the two polarization modes of the fiber have slightly
different propagation constants $\beta$. This birefringence
creates a refractive index difference of $\Delta n$ between the
polarization modes. In the fiber chosen in this experiment, the
holes form a hexagonal pattern that is slightly distorted to break
the hexagonal symmetry. Our fiber exhibits
strong birefringence $\Delta n$ of a few times $10^{-4}$. This
leads to non-negligible changes in the group velocity.

\begin{table}
  \centering
  \label{tab:fiber}
\begin{tabular}{|l|l|}
  \hline
  Property & Crystal Fiber $\backslash$  Erlangen   \\ \hline \hline
  Dispersion $D_0$ & $28\,\backslash 36 \,\mathrm{ps}/(\mathrm{nm}\,\mathrm{km})$ \\ \hline
  Dispersion $D_{m}$ & $-150 \,\backslash -180 \,\mathrm{ps}/(\mathrm{nm}\,\mathrm{km})$\\ \hline
  Third order dispersion $dD_{m}/d\lambda$ & $-0.6\,\backslash-0.75 \,\mathrm{ps}/(\mathrm{nm}^2\mathrm{km})$ \\ \hline
  Dispersion $\beta_2 (\lambda_0)$ & $-9.5\,\backslash-12\,\mathrm{ps}^2/\mathrm{km}$  \\ \hline
  Dispersion $\beta_2 (\lambda_\mathrm{m})$ & $180\,\backslash210\,\mathrm{ps}^2/\mathrm{km}$ \\ \hline
  Group velocity-matched wavelength $\lambda_\mathrm{m}$ & $1508\,\backslash1494\,\mathrm{nm} $\\ \hline
  Birefringence $\Delta n_0 \backslash \Delta n_\mathrm{m}$ & $7.5\,\backslash5.7\times10^{-4}$ \\ \hline
  Nonlinearity $\gamma$ ($780\,\mathrm{nm}$) & $0.1\,
  \mathrm{W}^{-1}\mathrm{m}^{-1}$ \\ \hline
  Fiber length $L$ & $1.5\,\mathrm{m}$ \\ \hline
\end{tabular}
  \caption[Fiber properties]{\small{Properties of fiber NL-PM-750B.
  Dispersion data according to Crystal Fiber
  $\backslash$ Alexander Podlipensky and
  Philip Russell, Max Planck Research Group in Optics, Information and     Photonics in Erlangen. Nonlinearity according to Crystal Fiber.
  The birefringence $\Delta n$ and the fiber length 
  $L$ were measured by  the authors. 
  The symbols used are defined in the text.}}
\end{table}

\subsection{Frequency shifts}

\begin{figure}[h]
\begin{center}
\includegraphics[width=30.0pc]{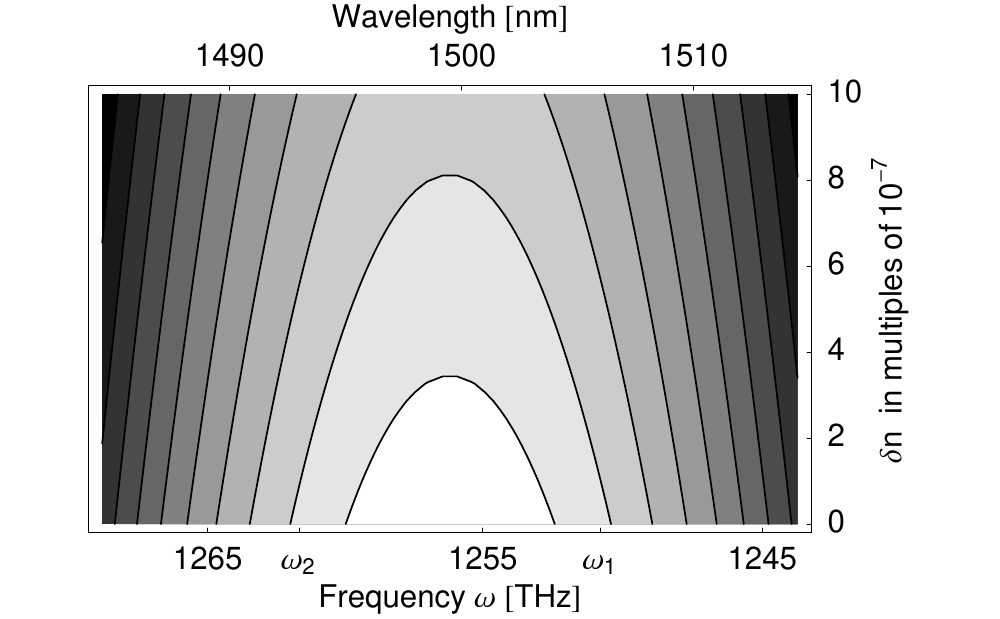}
\caption{
\small{
Doppler contours.
The pulse shifts the laboratory frequency $\omega$ 
(or the wavelength $\lambda$)
along the contour lines of
$\omega'$ as the function 
(\ref{eq:omegaprime0})
of the refractive-index change 
$\delta n=\chi/(2n_0)$.
For a sufficiently intense pulse 
$\delta n$ reaches the top of a contour.
In this case the probe light  
completes an arch on the diagram
while leaving the pulse;
it is red- or blue-shifted,
depending on its initial frequency.}
\label{fig:contours}}
\end{center}
\end{figure}

Consider the frequency shifts at a group velocity horizon.
During the pulse-probe interaction,
the co-moving frequency $\omega'$ is a conserved quantity 
and so 
the probe frequency $\omega$ follows a contour line of
$\omega'$ as a function of
the nonlinear susceptibility $\chi$ induced by the pulse,
see Fig.\ \ref{fig:contours}.
The maximal $\chi$ experienced by the probe is proportional to the
maximal nonlinear susceptibility $\chi_0$ experienced by the
pulse: assuming perfect mode overlap of pulse and probe,
$\chi_{\max}$ reaches $2\chi_0$ when the probe and the pulse are
co-polarized and $2\chi_0/3$ when they are cross-polarized, see
Sec.\ A1. If the pulse is a soliton, we obtain from Eq.\
(\ref{eq:soliton}) the peak susceptibility
%%%%%%%
\begin{equation}
\chi_0 = \frac{2n_0 c \lambda_0 D_0}{(\omega_0 T_0)^2}=\frac{2n_0
c |\beta_2(\lambda_0)| }{ \omega_0 T_0^2}
\end{equation}
%%%%%%%
where $D_0$ denotes the dispersion parameter at the carrier
wavelength $\lambda_0$. For example, for a soliton
(\ref{eq:soliton}) at $\lambda_0$=$800\,\mathrm{nm}$ whose full
width at half maximum (FWHM) is $70\,\mathrm{fs}$ (corresponding
to $T_0$=$40\,\mathrm{fs}$), for $n_0$=$1.5$,
$D_m$=$30\,\mathrm{ps}/(\mathrm{nm}\,\mathrm{km})$ the peak
susceptibility $\chi_0$ is as low as $2\times 10^{-6}$. 
Nevertheless, we show that this small variation in the optical
properties is sufficient to generate a significant wavelength
shift at the horizon.

%%%
%%%

We obtain the contours of $\omega'$ from the Doppler formula
(\ref{eq:doppler}). We use relations (\ref{eq:approxn}) and
(\ref{eq:betadispersion}), but integrate from the group
velocity-matching point,
%%%%%%%
\begin{eqnarray}
\omega' &=& \omega'_m - u \int_{\omega_m}^\omega
\int_{\omega_m}^\omega \beta_2\, d\omega\,d\omega -
\frac{\chi}{2n_0}\,\frac{u}{c}\,\omega
\nonumber\\
&\approx& \omega'_m - \frac{u}{2} \beta_2(\omega_m)\, 
(\omega-\omega_m)^2 - \frac{\chi}{2n_0}\,\frac{u}{c}\,\omega_m
\label{eq:omegaprime0}
\\
&\approx& \omega'_m + \frac{\pi u}{\lambda_m}
\left( \frac{D_m c
(\lambda-\lambda_m)^2}{\lambda_m} -
\frac{\chi}{n_0} \right) \,.
\label{eq:omegaprime}
\end{eqnarray}
%%%%%%%
The contours of $\omega'$ do not depend on $\omega_m'$
nor on the scaling factor ${\pi u}/{\lambda_m}$. Because
$D_m\,=\,D(\lambda_m)<0$, they form inverted
parabolas with a maximum at $\lambda_m$ for the
corresponding $\chi_{\max}$. They intersect the axis of zero
$\chi$ at the incident and the emerging wavelengths. Here
$|D_m| c (\lambda-\lambda_m)^2/\lambda_m$
equals $\chi_{\max}/n_0$, and so we get
%%%%%%%
\begin{equation}
\label{eq:lambdashift}
\lambda = \lambda_m \pm \delta\lambda \,,\quad
\delta\lambda =
\sqrt{\frac{\lambda_m\chi_{\max}}
{|D_m|\,n_0 c}}
\,.
\end{equation}
%%%%%%%
Using again that the pulse is a soliton, we obtain
%%%%%%%
\begin{equation}
\delta\lambda =
\frac{\sqrt{2r}\,\lambda_0}{T_0\sqrt{\omega_0\omega_m}}
\sqrt{\left|\frac{D_0}{D_m}\right|} =
\frac{\sqrt{2 r \lambda_m\lambda_0}}{T_0
\,\omega_m}
\sqrt{\left|
\frac{\beta_2(\lambda_0)}{\beta_2(\lambda_m)}
\right|}
\end{equation}
%%%%%%%
with $r$=$2$ for co-polarized and $r$=$2/3$ for cross-polarized
pulse and probe light. 
According to Fig.\ \ref{fig:contours} the probe
light can maximally be wavelength-shifted from $+\delta\lambda$ to
$-\delta\lambda$ over the range $2\delta\lambda$.
For the soliton mentioned above the group velocity dispersion
$D_0$ is about $30\,\mathrm{ps/(nm\,km)}$. Using
$\lambda_{\mathrm{m}}\approx 1500\,\mathrm{nm}$ and
$D_\mathrm{m}\approx-160\, \mathrm{ps/(nm\,km)}$, the wavelength
shift $2\delta\lambda$ is $20\,\mathrm{nm}$ in the co-polarized
case and $2\delta\lambda$=$12\,\mathrm{nm}$ in the cross-polarized
case.

We also derive a simple estimate of the efficiency of the
frequency shifting from the dispersion data. The probe
light that is colliding with the pulse undergoes frequency
conversion at the horizon. However, because the group velocities
of the probe $v_g$ and of the pulse $u$ are similar, only a small
fraction of the total probe light can be converted within the
finite length of the fiber. The pulse and the slightly faster
probe light travel through the fiber in $t$=$L/u$ and
$t_p$=$L/v_g$ with $t>t_p$. The time difference multiplied with
the probe power $P_\mathrm{probe}$ is the energy $E_\mathrm{coll}$
converted by pulse collision:
$E_\mathrm{coll}$=$P_\mathrm{probe}\,L\,(1/u-1/v_g)$. Therefore,
the fraction $\eta$ of probe power that is frequency converted is
%%%%%%%
\begin{equation}
\label{eq:efficiency}\eta =
\nu_\mathrm{rep}\,L\,(1/u-1/v_g)\,\approx
\nu_\mathrm{rep}\,L\,|D_m|\,\delta \lambda \,,
\end{equation}
%%%%%%%
where $\nu_\mathrm{rep}$ is the repetition rate of the pulses and
$1/v_g$=$\partial\beta/\partial\omega\approx 1/u +
D_m\delta\lambda$ was used. For $L$=$1.5\,\mathrm{m}$ and
$\nu_{\mathrm{rep}}$=$80\,\mathrm{MHz}$ 
the maximal conversion efficiency
$\eta$ is on the order of $10^{-4}$.

\subsection{Experimental results}

%%%
\begin{figure}[h]
\begin{center}
\includegraphics[width=27.0pc]{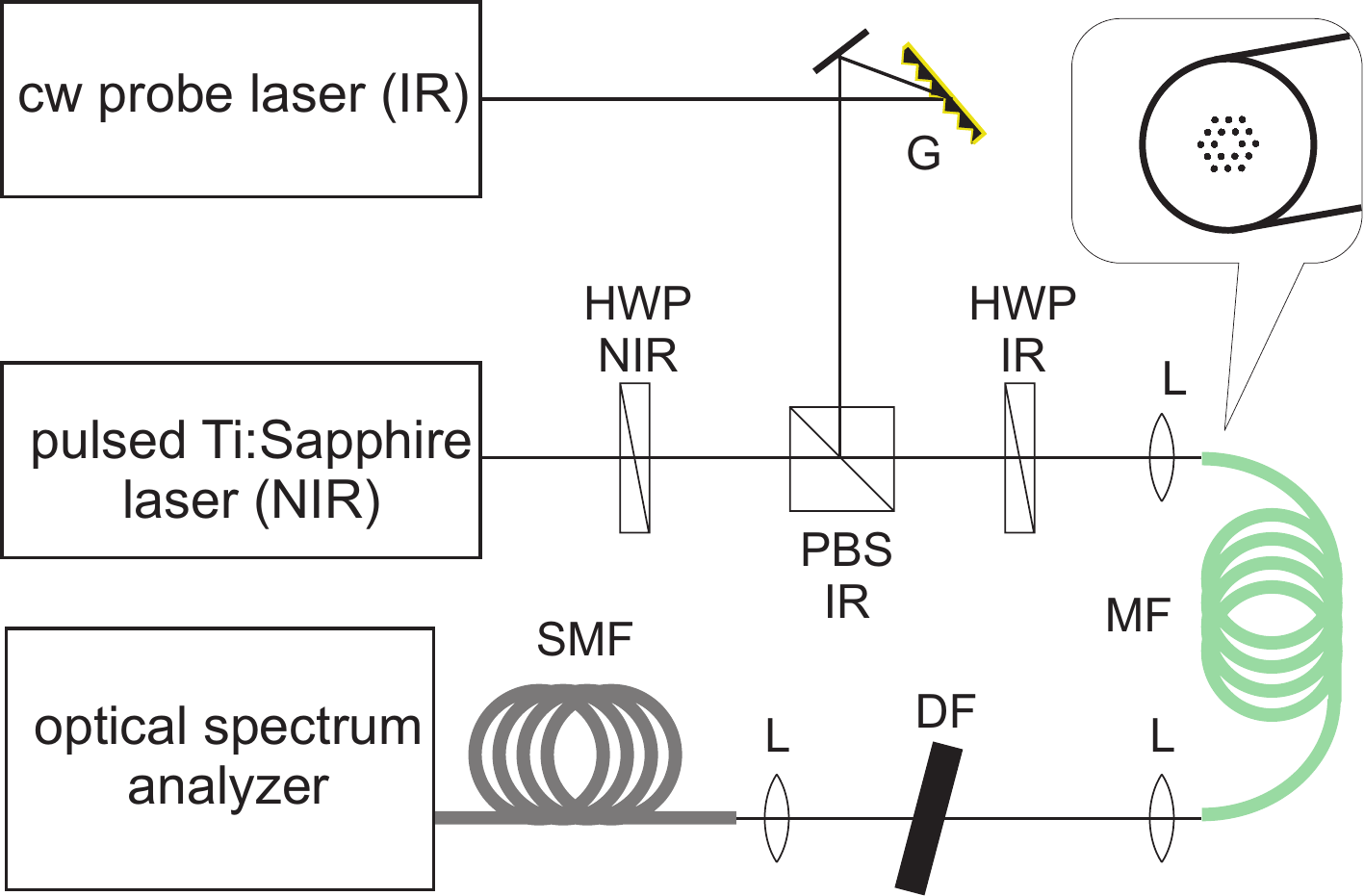}
\caption{ \small{Light from a continuous-wave infrared (IR) laser is
reflected off a diffraction grating (G) to suppress fluorescence.
The beam is steered into a microstructured fiber (MF) 
by a polarizing beam splitter (PBS),
a half wave plate (HWP), and a coupling lens (L). Near-infrared
(NIR) pulses are launched as well. After the fiber the pulses are
removed by a dichroic filter (DF) and the probe spectrum is taken
through a standard single mode fiber (SMF).} \label{fig:setup}}
\end{center}
\end{figure}
%%%

The experiment is arranged as displayed in Fig.\ \ref{fig:setup}.
A modelocked Ti:Sapphire laser (Mai Tai, Spectra Physics)
delivers 70-fs pulses (FWHM) in the near infrared (NIR). These
linearly polarized pulses are coupled to either one of the
principal axes of the microstructured fiber of length
$L$=$1.5\,\mathrm{m}$.  The polarization is rotated by a half-wave
plate. Note that the polarizing beam splitter (PBS) only acts on
the probe light. At the fiber output temporal autocorrelation
traces and spectra are taken to determine the pulse energy
necessary to create a fundamental soliton. For the center
wavelength of $803\,\mathrm{nm}$, a dispersion
$D_0$=$30\,\mathrm{ps}/(\mathrm{nm \,km})$ and a nonlinearity
$\gamma$ of $0.1\,\mathrm{W}^{-1}\mathrm{m}^{-1}$, 70-fs solitons
are generated at $5\,\mathrm{pJ}$ pulse energy corresponding to
$400\, \mathrm{\mu W}$ average power for the repetition rate
$\nu_\mathrm{rep}$=$80\,\mathrm{MHz}$.

The output pulse length equalled the 70-fs input pulse length at
an input power of approximately $320\, \mathrm{\mu W}$. This
indicates that a soliton has formed. 
The observed power of $320\, \mathrm{\mu W}$
in comparison with the predicted power of 
$400\, \mathrm{\mu W}$ illustrates the uncertainty
in the actual fiber dispersion and nonlinearity. 
The observed Raman-induced soliton self-frequency shift 
\cite{Agrawal,Mitschke} was 
$\lesssim 4\,\mathrm{nm}$. Note that this shift decelerates the pulse and
hence is changing the group velocity-matched wavelength
$\lambda_m$ in the infrared (IR). To calculate how much
$\lambda_m$ is shifted, we use Eq.\ (\ref{eq:gvddispersion}),
replacing $\lambda_0$ and $\lambda$ with
$\lambda_0+\delta\lambda_0$ and $\lambda_m+\delta\lambda_r$ and
linearize. In this way we get
%%%%%%%
\begin{equation}
\label{eq:raman}
\delta\lambda_r=
\frac{D_0}{D_m}\,\delta\lambda_0 \,.
\end{equation}
%%%%%%%
For the dispersion data shown in Fig.\ \ref{fig:dd}, a wavelength
change of $4\,\mathrm{nm}$ of the pulse changes $\lambda_m$ by
$\delta\lambda_r=-0.75\,\mathrm{nm}$. 
Since the probe light is wavelength-shifted
symmetrically around $\lambda_m$, there is a change of the
wavelength shift of up to $-1.5\,\mathrm{nm}$.

The probe light is derived from a tunable external grating diode
laser (Lynx Series, Sacher Lasertechnik). It delivers up to
$20\,\mathrm{mW}$ of continuous-wave light, tunable from $1460$ to
$1540\,\mathrm{nm}$. 
The probe light is reflected off a diffraction grating
to reduce fluorescence emitted near lasing bandwidth. 
With another half-wave plate the probe light is coupled
into the fiber onto one of the principal axes. Depending on
wavelength, $100$ to $600\, \mathrm{\mu W}$ of probe power were
coupled through the fiber. After the fiber we use a dichroic optic
to filter out all of the pulse light and couple the IR light into
a single-mode fiber connected to an optical spectrum analyzer.

\begin{figure}[t]
\begin{center}
\includegraphics[width=25.0pc]{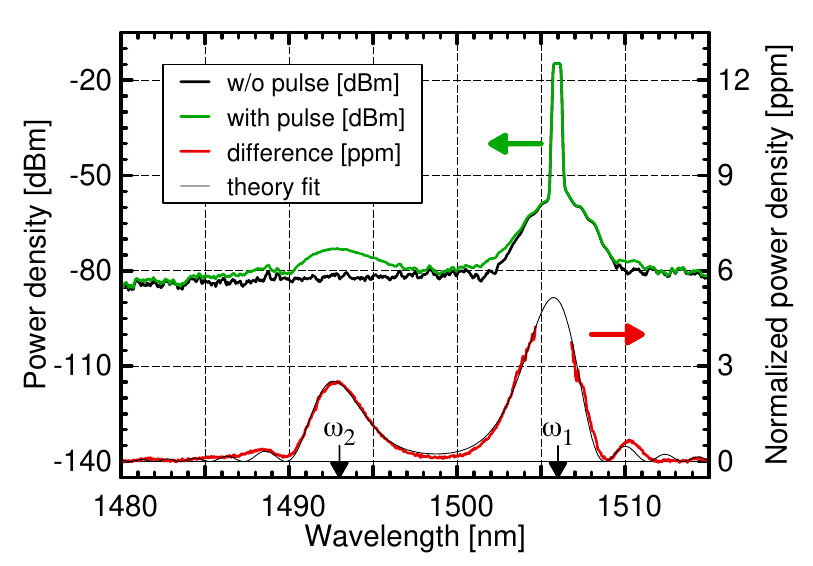}
\caption{ \small{Spectrum of the blue-shifted light
for initial probe of wavelength $1506\mathrm{nm}$.
Pulse and probe are co-polarized along the
slow axis of our fiber.
Traces with
(green) and without (black) probe light 
are shown on a logarithmic scale.
They are subtracted on a linear scale to obtain
the normalized signal (red) displayed in parts per million.}
\label{fig:spectrum}}
\end{center}
\end{figure}

Figure \ref{fig:spectrum} shows a typical output spectrum.
This spectrum was taken with pulse and probe aligned to the slow
axis of the fiber. At $\lambda$=$1506\,\mathrm{nm}$ the
diode-laser input line is visible as a strong signal. From
$\lambda$=$1502\,\mathrm{nm}$ 
to $\lambda$=$1510\,\mathrm{nm}$ we
detect residual weak spontaneous emission from the laser that was
not completely eliminated by the diffraction grating. 
Traces with
and without pulses present in the fiber are taken and subtracted,
leading to the signal displayed on a linear scale (red color). 
The signal is normalized by the amount of probe
power and by the resolution bandwidth of $0.5\,\mathrm{nm}$
and is given in parts per million (ppm). 
With the pulses present, a clear
peak appears on the blue side of the input probe light
near $1493\,\mathrm{nm}$.
Since the blue-shifted light is generated from the part of the probe
light that overlapped with the pulse during fiber propagation, it
constitutes itself a pulse of finite length. Hence, this length is
determined by the relative group velocity of probe light and the
pulse, see for example Eq.\ (\ref{eq:efficiency}). In turn, the unshifted
probe light is partially depleted, forming a gap in intensity.
These features lead to a spectral broadening of both the shifted and
unshifted probe light by a few nanometers.

%%%
%%%

From the measurements shown in Fig.\ \ref{fig:spectrum},
the efficiency of the blue-shifting is $1.1\times 10^{-5}$,
less than the estimated $10^{-4}$.
This indicates that a significant part of the probe 
light tunnels through the pulse;
the pulse is too short to establish a nearly perfect barrier. 
In the tunneling region of the pulse 
the laboratory frequency $\omega$ is purely imaginary.
In order to estimate the maximal imaginary part of $\omega$
we consider the extreme case where the initial frequency 
of the probe  
reaches the group-velocity-matched frequency $\omega_m$
characterized by
$\omega'=\omega_m'$.
We solve Eq.\  (\ref{eq:omegaprime0}) for $\omega$
and obtain
%%%%%%%
\begin{equation}
\mathrm{Im}\omega = 
\sqrt{\frac{\chi \omega_m}{n_0\beta_2(\omega_m) c}} =
\frac{\sqrt{\chi}\, \omega_m}
{\sqrt{n_0 c \lambda_m |D_m|}}
\,.
\end{equation}
%%%%%%%
Assuming $\chi\approx2\times 10^{-6}$ at the soliton peak,
$n_0\approx1.5$, $\lambda_m=1500\,\mathrm{nm}$ and
$D_m=-160 \,\mathrm{ps/(nm\,km)}$
the imaginary part of 
$\omega$ reaches
about $5\,\mathrm{THz}$.
This is insufficient to significantly suppress
tunneling through a 70-fs pulse,
because the product of $\mathrm{Im}\omega$
and $T_0$ is much smaller than unity.
For longer pulses we would expect perfect
frequency conversion at the horizon.

\begin{figure}[h]
\begin{center}
\includegraphics[width=25.0pc]{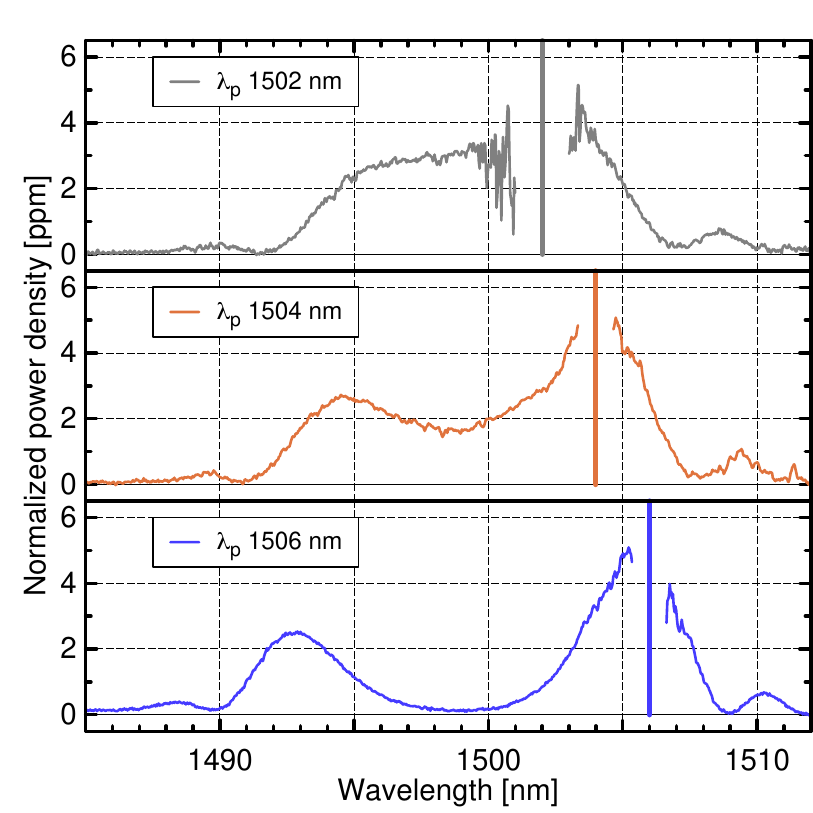}
\caption{ \small{ Spectra for different input probe wavelengths.
Since the probe mode is mirrored around 
the group-velocity-matched wavelength $\lambda_m$, increasing
probe wavelengths experience increasing blue shifting,
as is also illustrated by 
the contours of Fig.\ \ref{fig:contours}}.
\label{fig:tune}}
\end{center}
\end{figure}

Increasing the probe wavelength further away from $\lambda_m$ is
shifting light further to the blue side of the spectrum, because
the wavelength shifts symmetrically around the group
velocity-matched wavelength, according to 
Eq.\ (\ref{eq:lambdashift}) and Fig.\ \ref{fig:contours}.
Figure \ref{fig:tune} displays the spectra of shifted light for
three detunings of the probe light from the group velocity-matched
wavelength $\lambda_m$. As expected, the spectra move towards
shorter wavelengths by the same amount as the probe laser was
tuned towards longer wavelengths.

%%%
%%%

\begin{figure}[t]
\begin{center}
\includegraphics[width=25.0pc]{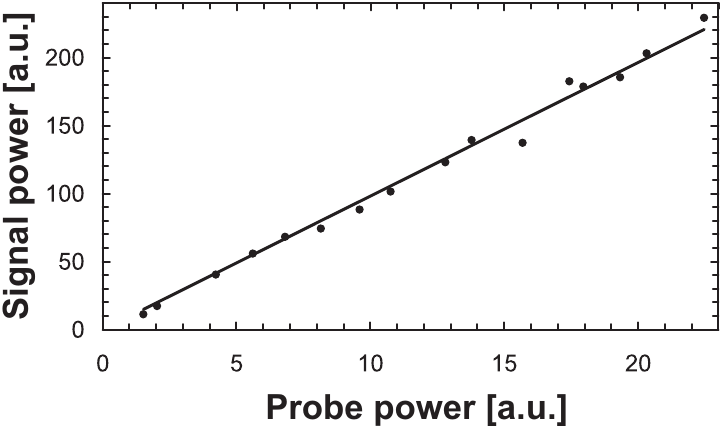}
\caption{ \small{Integrated blue-shifted spectrum versus incident
probe power. The solid line is a fit assuming proportionality.}
\label{fig:linearity}}
\end{center}
\end{figure}

Figure \ref{fig:linearity} shows how the signal strength, the
spectrum integrated over the signal peak, evolves with increasing
probe power. A clear linear dependence is evident in agreement
with our theoretical model and the superposition principle for the
probe light. The figure illustrates that the probe indeed is a probe, 
not influencing the pulses via nonlinear effects.

%%%
%%%

Changing the input polarizations changes the group velocities of
pulse and probe and therefore the group velocity-matched
wavelength $\lambda_m$ shifts by an amount $\delta \lambda_m$. If
we change for example the pump polarization from the fast to the
slow axis, the inverse group velocity $\beta_1(\lambda_0)$
increases as $n_{g0}/c$ is replaced by $(n_{g0}+\Delta n_0)/c$. To
maintain group velocity matching, $\beta_1(\lambda_m)$ has to
change accordingly by $\Delta n_0/c$. We use Eq.\
(\ref{eq:gvddispersion}), linearizing around $\lambda_m$, and get
%%%%%%%
\begin{equation}
\Delta n_0/c \approx D_m \delta\lambda_m \,.
\end{equation}
%%%%%%%
For $D_m$=$-160\, \mathrm{ps/(nm\,km)}$ and $\Delta
n_0$=$7.5\times10^{-4}$ we obtain  $\delta\lambda_m \!\approx\!
-16\,\mathrm{nm}\approx 2\delta\lambda$. This means that, when
changing polarizations, the probe laser has to be retuned to a
wavelength were frequency shifting can be observed.

\begin{figure}[h]
\begin{center}
\includegraphics[width=35.0pc]{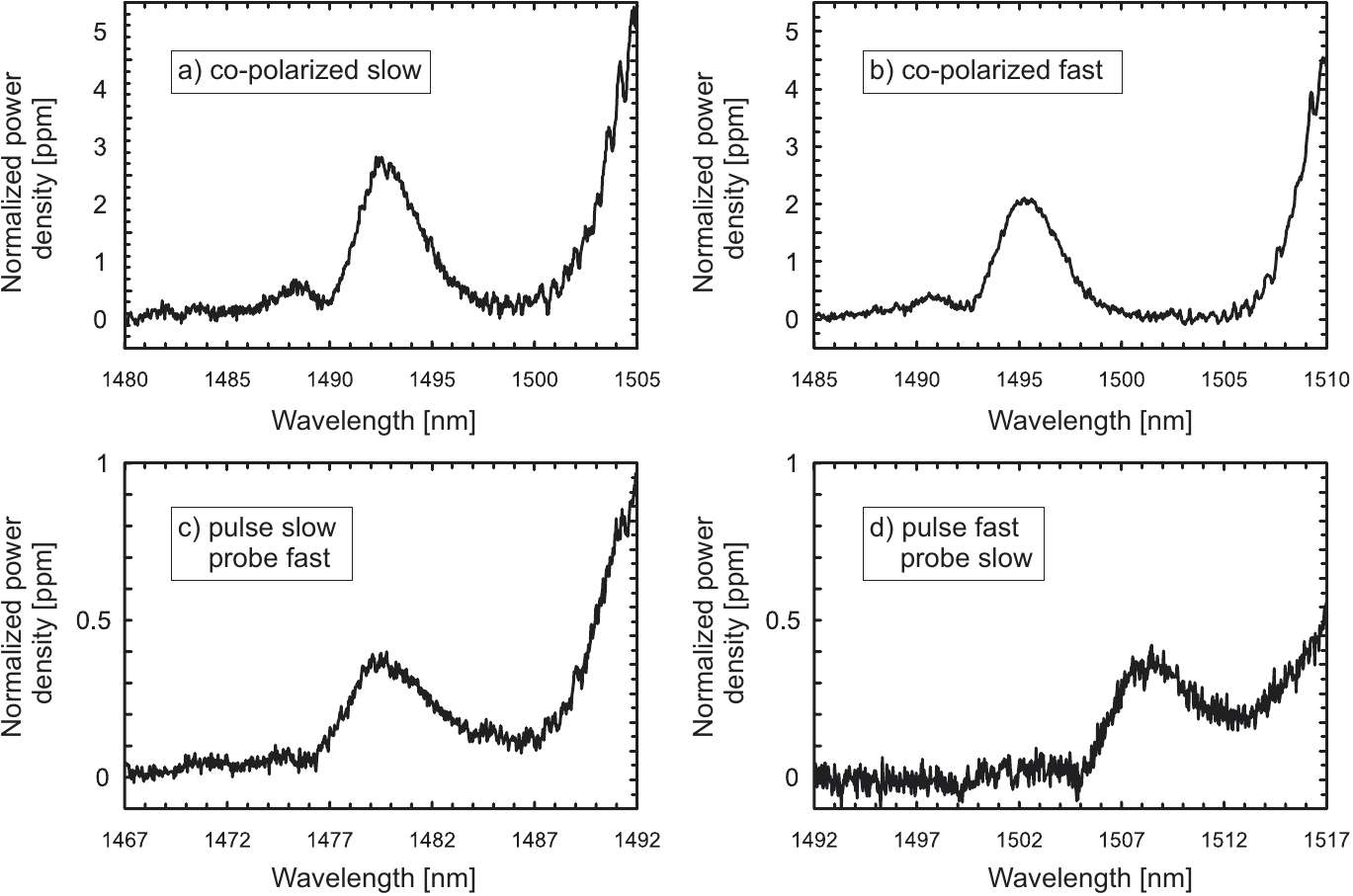}
\caption{ \small{ Blue-shifted spectra for all four polarization
combinations. Co-polarized spectra on the slow and fast axis in
(a) and (b) and cross-polarized spectra with pulses slow (c) and
fast (d). Group velocity-matched wavelengths are (a)
$1499.5\,\mathrm{nm}$ (b) $1503.2\,\mathrm{nm}$ (c)
$1486.4\,\mathrm{nm}$ and (d) $1513.3\,\mathrm{nm}$. }
\label{fig:pol}}
\end{center}
\end{figure}

Figure \ref{fig:pol} shows spectra for all four different
polarization combinations. As expected, the group velocity-matched
wavelength changes. Note that there also is a difference in
$\lambda_m$ for the two co-polarized cases, indicating small
changes in the dispersion profile for the two polarization axes, a
dispersion of the birefringence.

%%%
%%%

\subsection{Calculation of spectra}

Here we derive a functional expression for the spectra of the
frequency-shifted probe light, applying the theory developed in
Appendix A. 
We focus on the part of the incident light that interacts 
with the pulse and ignore the component that tunnels through,
because the tunneled light does not contribute
to the observed spectrum.
We represent the relevant probe light $A$ as a superposition
(\ref{eq:modes}) of stationary modes $A_R$ and $A_L$ on the right
or left side of the horizon that are characterized by the
frequency $\omega'$ in the co-moving frame. We replace the mode
operators by classical amplitudes $a_R$ and $a_L$ and focus on the
positive-frequency component of $A$. In this way we obtain
%%%%%%%
\begin{equation}
\label{eq:rlexp}
A = \int (a_R A_R + a_L A_L )\, d\omega' \,.
\end{equation}
%%%%%%%
For the mode functions $A_R$ and $A_L$ we use the theory of
geometrical optics in moving media developed in Sec.\ A.8. We
assume that the pulse is infinitely short in comparison with
variations of the probe light; and write down the modes as
%%%%%%%
\begin{eqnarray}
\label{eq:rlmodes}
A_R &=& \Theta(\tau)\left({\cal A}_1e^{-i\omega_1\tau} +i {\cal
A}_2 e^{-i\omega_2\tau}\right) e^{-i\omega'\zeta} \,,
\nonumber\\
A_L &=& \Theta(-\tau)\left({\cal A}_1e^{-i\omega_1\tau} -i {\cal
A}_2 e^{-i\omega_2\tau}\right) e^{-i\omega'\zeta} \,.
\end{eqnarray}
%%%%%%%
Here $\omega_1$ denotes the laboratory frequency with respect to
$\omega'$ and  $\omega_2$ the blue-shifted laboratory frequency.
The factors of $\pm i$ describe the $\pi/2$ phase shifts at
turning points \cite{LL3}. 
Their sign depends on the sign of the frequency change
and the side of the horizon.  
We assume that $\omega_1$ and
$\omega_2$ are sufficiently close to the group velocity-matched
frequency $\omega_m$
such that we can use the quadratic approximation
(\ref{eq:omegaprime0}) for $\omega'$ 
outside the pulse where $\chi$ vanishes,
%%%%%%%
\begin{equation}
\omega' = 
 \omega'_m - \frac{u}{2} \beta_2\, 
(\omega-\omega_m)^2  \,,
\label{eq:omegaprime2}
\end{equation}
%%%%%%%
which implies that the two solutions $\omega_1$ and $\omega_2$ for
a given $\omega'$ are symmetric around the group velocity-matched
frequency $\omega_m$, such that
%%%%%%%
\begin{equation}
\label{eq:omega2}
\omega_2  =
\omega_m+(\omega_m-\omega_1)  =
2\omega_m - \omega_1
\,.
\end{equation}
%%%%%%%
The amplitudes ${\cal A}_1$ and ${\cal A}_2$ are given by the
expression (\ref{eq:amplitude}). Since $n\omega$ does not vary
much with frequency and $\partial\omega_1/\partial\omega'
$=$-\partial\omega_2/\partial\omega'$, we approximate
%%%%%%%
\begin{equation}
\label{eq:amps}
{\cal A}_1 \approx {\cal A}_2 =
{\cal A}_0 \left|\frac{\partial\omega_1}{\partial\omega'}\right|
\,.
\end{equation}
%%%%%%%
Finally,
we write the mode expansion (\ref{eq:rlexp})
as an infinite integral over $\omega_1$,
assuming that contributions outside the physically
relevant range of $\omega_1$ are negligible,
%%%%%%%
\begin{equation}
\label{eq:modeintegral}
A = \int_{-\infty}^{+\infty}
\big(a_R(\omega_1) A_R + a_L(\omega_1) A_L \big)\,
d\omega_1
\,.
\end{equation}
%%%%%%%
Formulas 
(\ref{eq:rlmodes}-\ref{eq:modeintegral})
specify our theoretical model.

In our experiment, we measure the modulus squared of the Fourier
transform of $A$ with respect to the laboratory time $t$ at the
end of the fiber $z$=$L$. This is identical to the modulus squared
of the Fourier transform $\widetilde{A}$ with respect to the
retarded time $\tau$=$t-z/u$ at $\zeta$=$L/u$. Using the standard
relations
%%%%%%%
\begin{equation}
\label{eq:halfdelta} \int_0^{+\infty} e^{i\omega\tau} d\tau =
\frac{i}{\omega} + \pi\delta(\omega) \,,\quad \int_{-\infty}^0
e^{i\omega\tau} d\tau = -\frac{i}{\omega} + \pi\delta(\omega)
\end{equation}
%%%%%%%
we obtain the result
%%%%%%%
\begin{eqnarray}
\widetilde{A} 
&=&
\frac{1}{2\pi}
\int_{-\infty}^{+\infty} A(\tau) e^{i\omega \tau}\, d\tau
\nonumber\\
&=&
{\cal A}_0\, \frac{a_R(\omega)+a_L(\omega)}{2}
e^{-i\omega'\zeta}
 +i {\cal A}_0\,
 \frac{a_R(2\omega_m - \omega)-a_L(2\omega_m - \omega)}{2}
e^{-i\omega'\zeta}
\nonumber\\
&& + \int_{-\infty}^{+\infty}
{\cal A}_0\, \bigg(
\frac{ia_R(\omega_1)-i a_L(\omega_1)}
{2\pi(\omega-\omega_1)} -
\frac{a_R(\omega_1)+a_L(\omega_1)}
{2\pi(\omega-\omega_2)} \bigg)
e^{-i\omega'\zeta}\,
d\omega_1
\label{eq:atilde}
\end{eqnarray}
%%%%%%%
where the integral is understood as a Principal Value Integral
\cite{Ablowitz} through the poles where $\omega_1$ or $\omega_2$
go through $\omega$. In the first line of Eq.\ (\ref{eq:atilde})
$\omega'$ is a function of $\omega$ and in the integral in the
second line $\omega'$ is understood to be a function of
$\omega_1$. 

At the entrance of the fiber, the initial probe $A_p$ is a
continuous wave with frequency $\omega_p$ and an amplitude we
denote as $2{\cal A}_0$. The coefficients $a_R$ and $a_L$ that
describe this situation are given by the expressions
%%%%%%%
\begin{equation}
\label{eq:coeff}
a_R = \delta(\omega_1-\omega_p)
+ \frac{i}{\pi(\omega_1-\omega_p)}
\,,\quad
a_L = \delta(\omega_1-\omega_p)
- \frac{i}{\pi(\omega_1-\omega_p)}
\,,
\end{equation}
%%%%%%%
as one verifies by the following procedure: we substitute the
coefficients in the integral (\ref{eq:modeintegral}), extract the
contribution of the delta function and apply Cauchy's Residue
Theorem \cite{Ablowitz} for the remaining integral. For this, we
close the integration contours on the complex half planes where
the integrand exponentially decreases. For the modes on the
right-hand side of the horizon, where $\tau>0$, we chose the lower
half plane for the $\exp(-i\omega_1\tau)$ term and the upper half
plane for $\exp(-i\omega_2\tau)$, in view of the relationship
(\ref{eq:omega2}); on the left-hand side we take the opposite
planes. Since we integrate through the poles we obtain half of the
residue, similar to the derivation of Hilbert transformations
\cite{Ablowitz} or Kramers-Kronig relations  \cite{Toll}. The
result of this calculation is the incident plane wave $2{\cal
A}_0\exp(-i\omega_p\tau)$, which justifies the mode coefficients
(\ref{eq:coeff}).

In order to calculate the spectrum,
we substitute the mode coefficients (\ref{eq:coeff})
into Eq.\ (\ref{eq:atilde}).
The delta functions immediately generate
contributions to the spectrum;
we focus on the calculation of
the remaining Principal Value Integral
through $(\omega-\omega_1)^{-1}(\omega_1-\omega_p)^{-1}$
with Gaussian $\exp(-i\omega'\zeta)$
given by Eq.\ (\ref{eq:omegaprime2}).
We represent
$(\omega-\omega_p)
(\omega-\omega_1)^{-1}(\omega_1-\omega_p)^{-1}$
as $(\omega_1-\omega_p)^{-1}-(\omega_1-\omega)^{-1}$
and use the Hilbert transform of a Gaussian
\cite{LMKRR},
%%%%%%%
\begin{equation}
G(x) =
\int_{-\infty}^{+\infty}
\frac{e^{-\xi^2}d\xi}{\pi(x-\xi)}
= e^{-x^2}\mathrm{erfi}(x)
\,,\quad
\mathrm{erfi}(x) = \frac{2}{\sqrt{\pi}}
\int_0^x e^{\xi^2} d\xi
\,.
\end{equation}
%%%%%%%
In this way we find that the spectral field $\widetilde{A}$
consists of the sum of the delta peak ${\cal A}_0
\delta(\omega-\omega_p)\, \exp(-i\omega_p'\zeta)$ and the
contribution that describes the frequency shifting of the probe,
%%%%%%%
\begin{eqnarray}
\widetilde{A}_s
& = &  \frac{{\cal A}_0}
{\pi(\omega+\omega_p-2\omega_m)}
\Big[\exp\left(iq(\omega-\omega_m)^2\right)-
\exp\left(iq(\omega_p-\omega_m)^2\right)
\Big]
\nonumber\\
&&
- \frac{{\cal A}_0}
{\pi(\omega-\omega_p)} \,
\left[G\left(\sqrt{-iq}\,(\omega-\omega_m)\right)
- G\left(\sqrt{-iq}\,(\omega_p-\omega_m)\right)
\right]
\label{eq:spectrum0}
\end{eqnarray}
%%%%%%%
where we ignored the unimportant overall phase of
$\omega_m'\zeta$
and used the abbreviation
%%%%%%%
\begin{equation}
q = \frac{\pi c |D_m| L}{\omega_m^2}
\,.
\end{equation}
%%%%%%%
So, up to an overall phase, the spectral field
$\widetilde{A}$ is given in terms of experimentally accessible
parameters, the fiber length $L$, the dispersion $D_m$ and the
group velocity-matched frequency $\omega_m$.

\begin{figure}[h]
\begin{center}
\includegraphics[width=25.0pc]{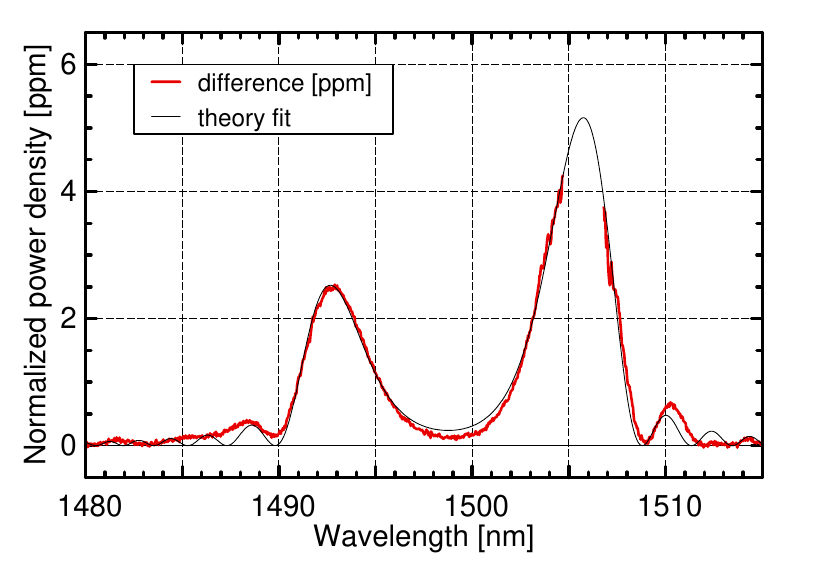}
\caption{ \small{Fit with theory.
The red curve shows the normalized difference data
of Fig.\ \ref{fig:spectrum}
and the thin black line the fit with 
the theoretical expression (\ref{eq:spectrum}).}
\label{fig:fit}}
\end{center}
\end{figure}

As expected, the spectrum gets narrower around the blue-shifted
and probe frequencies $2\omega_m-\omega_p$ and $\omega_p$ with
increasing propagation distance in the fiber, because both the
converted and depleted components of the probe light form longer
pulses for longer interaction times. However, the carrier
frequency of the pulse is gradually red-shifted due to the soliton
self-frequency shift (SFS) \cite{Mitschke}.
According to Eq.\ (\ref{eq:raman}), 
this leads to a shift $\delta\lambda_r$ in the group
velocity-matched wavelength $\lambda_m$ of about
$-0.75\,\mathrm{nm}$ along the fiber. The blue-shifted light is
created at decreasing wavelengths
and also the part of the pulse that interacts with the probe 
adiabatically follows.
In the signal spectrum, 
both the group-velocity-matched frequency $\omega_m$
and the probe frequency $\omega_p$
appear to be shifted
by $\delta\omega_r = -(\delta\lambda_r/\lambda_m)\omega_m$
and is replaced by $\omega_s$.
Since initially the spectrum is broad,
we incorporate the SFS effect in a phenomenological form in
our formula by a reduced efficiency $\eta_r$ for the blue-shifted
part of the spectrum as
%%%%%%%
\begin{eqnarray}
\widetilde{A}_s
& = &  \frac{\eta_r {\cal A}_0}
{\pi(\omega+\omega_s-2\omega_m)}
\Big[\exp\left(iq(\omega-\omega_m)^2\right)-
\exp\left(iq(\omega_s-\omega_m)^2\right)
\Big]
\nonumber\\
&&
- \frac{{\cal A}_0}
{\pi(\omega-\omega_s)} \,
\left[G\left(\sqrt{-iq}\,(\omega-\omega_m)\right)
- G\left(\sqrt{-iq}\,(\omega_s-\omega_m)\right)
\right]
\,.
\label{eq:spectrum}
\end{eqnarray}
%%%%%%%
Figure \ref{fig:fit} shows the fit
of the observed spectrum of Fig.\ \ref{fig:spectrum}
with the theoretical curve (\ref{eq:spectrum}). As fitting
parameters we used the overall amplitude ${\cal A}_0$, the shifted 
probe frequency $\omega_s$ and group-velocity-matched
frequency $\omega_m$ (in terms of the corresponding 
wavelengths), the dispersion $D_m$ and $\eta_r$. 
We obtain a very good fit for
${\cal A}_0=4.1\times10^{12}\sqrt{\mathrm{W}}\mathrm{s}$,
$\lambda_s=1505.31\, \mathrm{nm}$,
$\lambda_m=1499.38\, \mathrm{nm}$,
$D_m=-187 \mathrm{ps/(nm\,km)}$
and $\eta_r=0.80$.
The shift in $\lambda_p$ is consistent 
with the effect (\ref{eq:raman}) of the soliton
self-frequency shift. 
The fitted values for $D_m$ and $\lambda_m$ agree with
the independently measured dispersion and the 
group-velocity-matched frequency 
calculated from the dispersion curve of Fig.\ \ref{fig:dd}.

In conclusion, we have shown that light was blue-shifted by a near
group velocity-matched pulse. The measured data was explained by
the presence of an optical group velocity horizon inside the
fiber. A very good agreement between theory and
experiment was achieved. The blue shifting corresponds to the
optical analogue of trans-Planckian frequency shifts in
astrophysics \cite{TransPlanck}. In this way, we have demonstrated
classical optical effects of the event horizon in our analogue
system, a first step towards tabletop astrophysics \cite{Ball}.

%%%
%%%

\end{appendix}


\begin{thebibliography}{99}
\bibitem{Remark_g}
The exact wave equation in a curved space time geometry is
$\partial_\mu \sqrt{-g} g^{\mu\nu}\partial_\nu A $=$ 0$. where $g$
is the determinant of the metric tensor \cite{LL2}. In the case
(\ref{eq:gmunu}) $g$ depends only on the refractive index $n$ and
hence $g$ is constant for constant $n$.
\end{thebibliography}
\end{document}